%% file: A_main.tex
\documentclass[sigconf]{acmart}

\usepackage{natbib}
\usepackage{times}
\usepackage{soul}
\usepackage{hyperref}
\usepackage[utf8]{inputenc}
\usepackage{caption}
\usepackage{graphicx}
\usepackage{amsmath}
\usepackage{amsthm}
\usepackage{booktabs}
\usepackage{algorithm}
\usepackage{algorithmic}
\usepackage[switch]{lineno}

\usepackage{adjustbox}
\usepackage{multirow}
\usepackage{multicol}   

\usepackage{newfloat}
\usepackage{listings}
\usepackage{xcolor}

\AtBeginDocument{%
  }

\title{CHAT: Beyond Contrastive Graph Transformer for Link Prediction in Heterogeneous Networks}
\begin{document}

\author{Shengming Zhang}
\email{michaelzhang@ibms.pumc.edu.cn}
\orcid{0009-0001-4549-3262}
\affiliation{%
  \institution{Chinese Academy of Medical Sciences \& Peking Union Medical College}
  \city{Beijing}
  \country{China}
}
\author{Le Zhang}
\email{zhangle09@baidu.com}
\orcid{0000-0003-0894-9651}
\affiliation{%
  \institution{Baidu Research}
  \city{Beijing}
  \country{China}
}

\author{Jingbo Zhou}
\email{zhoujingbo@baidu.com}
\orcid{0000-0003-2677-7021}
\affiliation{%
  \institution{Baidu Research}
  \city{Beijing}
  \country{China}
}

\author{Hui Xiong}
\email{xionghui@hkust-gz.edu.cn}
\orcid{0000-0001-6016-6465}
\affiliation{%
  \institution{Hong Kong University of Science and Technology (Guangzhou)}
  \city{Guangzhou}
  \country{China}
}





\begin{abstract}
Link prediction in heterogeneous networks is crucial for understanding the intricacies of network structures and forecasting their future developments. Traditional methodologies often face significant obstacles, including over-smoothing—wherein the excessive aggregation of node features leads to the loss of critical structural details—and a dependency on human-defined meta-paths, which necessitate extensive domain knowledge and can be inherently restrictive. These limitations hinder the effective prediction and analysis of complex heterogeneous networks.
In response to these challenges, we propose the \underline{C}ontrastive \underline{H}eterogeneous gr\underline{A}ph \underline{T}ransformer (\textbf{CHAT}). \textbf{CHAT} introduces a novel sampling-based graph transformer technique that selectively retains nodes of interest, thereby obviating the need for predefined meta-paths. The method employs an innovative connection-aware transformer to encode node sequences and their interconnections with high fidelity, guided by a dual-faceted loss function specifically designed for heterogeneous network link prediction. Additionally, \textbf{CHAT} incorporates an ensemble link predictor that synthesizes multiple samplings to achieve enhanced prediction accuracy.
We conducted comprehensive evaluations of \textbf{CHAT} using three distinct drug-target interaction (DTI) datasets. The empirical results underscore \textbf{CHAT}'s superior performance, outperforming both general-task approaches and models specialized in DTI prediction. These findings substantiate the efficacy of \textbf{CHAT} in addressing the complex problem of link prediction in heterogeneous networks.
\end{abstract}

\maketitle

\input{LaTeX/abstract_introduction}

\input{LaTeX/related_work}

\input{LaTeX/preliminary}

\input{LaTeX/methodology}

\input{LaTeX/Experiments}

\bibliographystyle{ACM-Reference-Format}
\bibliography{reference}

\include{LaTeX/appendix}


\end{document}

%% file: LaTeX/abstract_introduction.tex
\section{Introduction}

Networks provide a versatile framework for representing intricate relationships and interactions across diverse domains~\citep{borgatti2009network}.
Various forms of networks are prevalent across multiple domains, each serving as a distinctive framework of interactions,
e.g. relationships within social networks~\citep{marin2011social}, chemical and biological interactions~\citep{udrescu2016clustering,luo2017network}, academia citation networks~\citep{dong2017metapath2vec}, and investment networks in entrepreneurship~\citep{zhang2021scalable}.
Link prediction seeks to estimate the likelihood of interaction existence between two nodes based on observed links and node attributes~\citep{kumar2020link}.
In biological networks, it may contribute to predicting unexplored drug-target interactions with implications for new drugs~\citep{luo2017network}. Thus, the study of network evolution and link prediction underscores the significance of network analysis across various domains. 

\begin{figure}[t]
    \centering

    \includegraphics[width = 1\columnwidth]{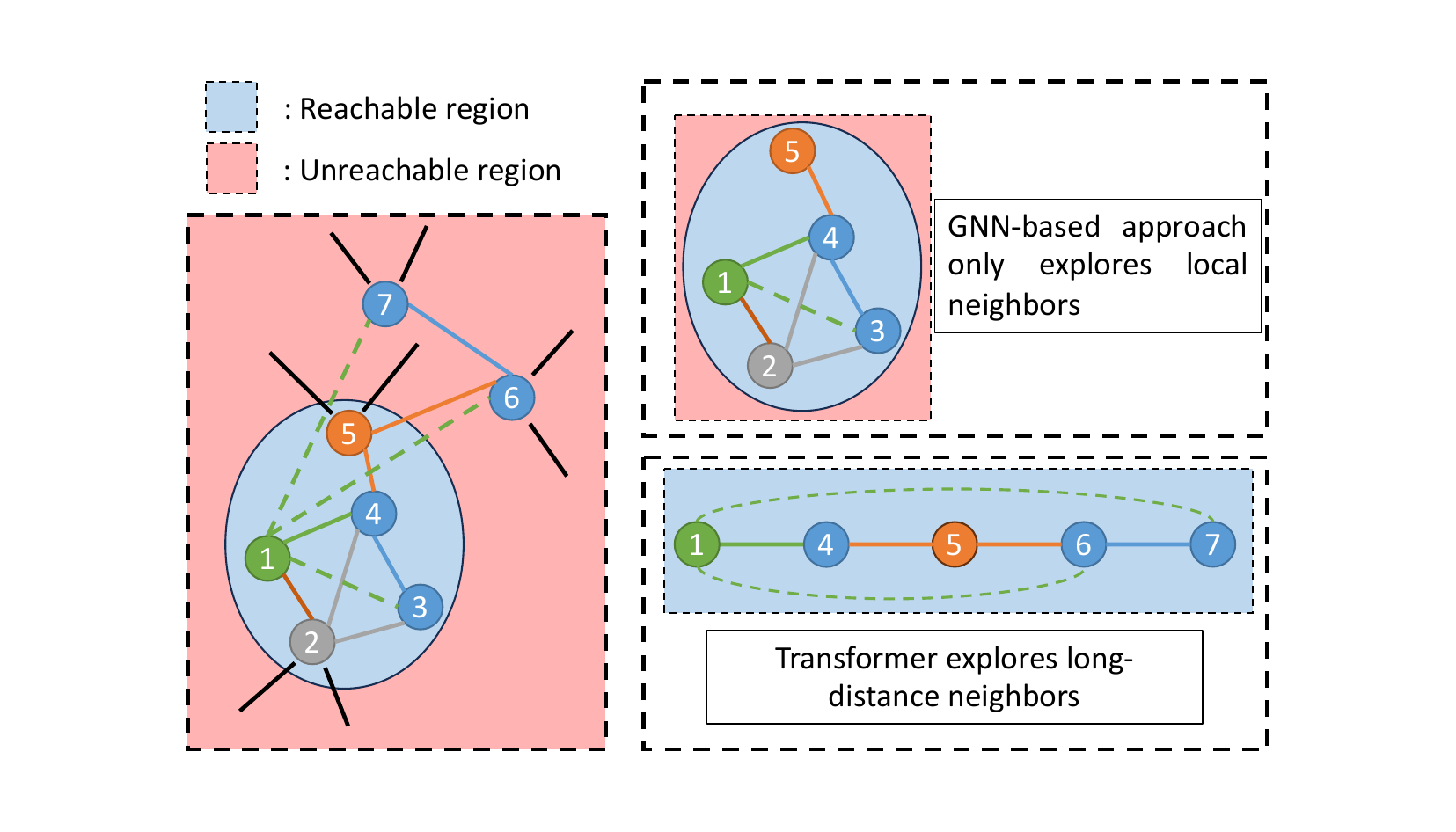}
    \caption{Comparison between GNN and transformer-based approaches in heterogeneous networks. GNN explores up to k-hop neighbors, where nodes beyond k-hop are unreachable (Red region), increasing k leads to over-smoothing issues. Transformer explores neighbors of long-distance, extending the reachable region (blue region) without over-smoothing.}
    \vspace{-2mm}
    \label{fig:cover}
\end{figure}
In the past few decades, substantial scholarly efforts have been dedicated to addressing the issue of link prediction in network analysis. Broadly, these methodologies can be classified into three categories: (i) Similarity-based approaches compute edge scores based on similarity measures, such as Jaccard similarity~\citep{jaccard1901etude} and cosine similarity~\citep{salton1983introduction}. Despite their simplicity and intuitiveness, these approaches are largely dependent on the features extracted from nodes. They may inadvertently overlook existing interactions within the network, which inherently carry rich information. (ii) Probabilistic models build a statistical framework with potential edges as variables using a limited set of model parameters, and maximize the conditional probability, with examples including Probabilistic Relational Models (PRM)~\citep{getoor2001learning} and Probabilistic Soft Logic Models (PSL)~\citep{bach2017hinge}. However, these models necessitate solving an inference problem over the entire network, which results in high computational cost in terms of time and space. Additionally, the limited number of parameters restricts the expressive power of such models, potentially impairing their prediction performance. (iii) Machine Learning-based approaches, particularly deep learning methodologies, leverage Graph Neural Networks (GNNs) to learn node representations and unearth novel connections within the representation space~\citep{cai2021line,zhu2021neural}. Despite their demonstrated effectiveness, these approaches face several challenges, including over-smoothing, domain-specific knowledge reliance, and scalability issues.

Transformer models have witnessed considerable success in handling problems related to sequential data~\citep{zhang2022cat,devlin2018bert}, with burgeoning attempts to adapt these models for graph data~\citep{ying2021transformers,zhang2022hierarchical}. One of the benefits of utilizing transformer-based models in graph data is the avoidance of over-smoothing issue due to the effective self-attention mechanism~\citep{muller2023attending}. Figure~\ref{fig:cover} illustrates a comparison between GNN-based approach and transformer-based approach. 

Despite the advantages of transformer-based approaches, applying existing transformer-based models directly for link prediction in heterogeneous networks presents a triad of challenges:
(i) Existing transformer-based models predominantly concentrate on small graphs, such as in chemical compound representation learning~\citep{ying2021transformers}, limiting their applicability for large-scale networks.
(ii) The original design of transformer models overlook the connections between tokens, while it is crucial to consider in heterogeneous networks.
(iii) The primary focus of existing transformer-based graph models is the learning of graph or node representations~\citep{kreuzer2021rethinking,rampavsek2022recipe}, and their application does not naturally extend to link prediction tasks in heterogeneous networks.

To overcome the limitations inherent in existing transformer models, we introduce a novel sampling-based approach - the \underline{C}ontrastive \underline{H}eterogeneous gr\underline{A}ph \underline{T}ransformer (\textbf{CHAT}). \textbf{CHAT} is specifically tailored for link prediction task within heterogeneous networks. \textbf{CHAT} employs a concentrated graph random-walk sampling technique that selects nodes of interest from the heterogeneous network, subsequently generating sequences of graph samples.
The concentrated sampling scheme thresholds the sample size, moderating scalability issue.
Our proposed sampling technique also features at its ability to explore comprehensive connections without the requirement for pre-defined meta-paths.

Subsequent to the sampling process, a connection-aware transformer is utilized that encodes both nodes and connections across sampled sequences.
The 
connection-aware transformer is supervised by a dual-faceted loss function: a supervised contrastive link prediction loss that forces distinction between linked and unlinked nodes, and
an observation probability loss enforcing proximity between connected nodes. An ensemble link predictor is proposed to force agreements between samples. 
In order to examine the effectiveness of \textbf{CHAT}, we choose drug-target interaction (DTI) prediction as our targeted domain. Comprehensive experimental evaluations spanning three drug-target interaction prediction datasets demonstrate that \textbf{CHAT} consistently outperforms both conventional benchmarks and state-of-the-art DTI prediction approaches crafted with domain-specific knowledge. Distilling the essence of our work, we highlight three pivotal technical contributions encapsulated in this paper:
\begin{itemize}
    \item We propose \textbf{CHAT}, a novel framework in predicting potential links in heterogeneous networks.
    \item We develop a concentrated graph sampling technique that captures comprehensive connections over heterogeneous networks, skipping the need of pre-defined meta-paths and alleviating scalability challenges. 
    \item \textbf{CHAT} incorporates a connection-aware transformer that incorporates both connections and nodes, traversing long-distance node neighbors while adeptly circumventing the ubiquitous over-smoothing dilemma.
    \item \textbf{CHAT} equipts a novel dual-faceted loss function together with an ensemble link predictor in supervising the connection-aware transformer as well as predicting potential links over heterogeneous networks.
\end{itemize}


%% file: LaTeX/related_work.tex
\section{Related Works}

\subsection{Link Prediction in Heterogeneous Networks}
Link prediction in heterogeneous networks has gained significant attention in recent decades, becoming a vital field of study. This area focuses on predicting potential interactions between diverse entities. A common example is identifying possible new connections in social networks, such as predicting future friendships~\citep{huang2015social}. The scope of link prediction, however, extends much further, influencing various domains. Notable applications include forecasting drug-drug and drug-target interactions in the biology~\citep{lin2020kgnn,wen2017deep} networks, projecting venture capital investments in the entrepreneurial activity networks~\citep{xu2022sociolink}, and predicting click-through rates in the e-commerce user-item networks~\citep{zhou2019deep}.

A particularly notable area within link prediction is drug-target interaction (DTI) prediction. DTI prediction stands out due to its practical significance and the inherent heterogeneity of its networks. These networks often include not just primary nodes like drugs and targets but also ancillary nodes such as diseases and side effects. Therefore, DTI prediction serves as an exemplary model for studying link prediction dynamics in heterogeneous networks. 

\subsection{Link Prediction Approaches}

Researchers have explored various methods in heterogeneous network link prediction, e.g. similarity-based strategies~\citep{salton1983introduction} and probabilistic models~\citep{bach2017hinge}. Another approach is matrix factorization, which represents different types of nodes through latent vectors specific to each~\citep{liu2016neighborhood}. Additionally, network diffusion algorithms have been used for learning low-dimensional representations~\citep{luo2017network}.

Recently, Graph Neural Network (GNN)-based methods have become prominent in link prediction. These methods aim to learn effective node and link representations by capturing both topological and attribute information within the network. Specifically, \citet{zhang2020revisiting} incorporate an auto-encoder-based approach with a labeling trick for structural link prediction; \citet{cai2020multi} focus on multi-scale node aggregation over sampled subgraphs; \citet{zhu2021neural} adopt path formulation and a generalized neural Bellman-Ford algorithm for edge representation learning. 

There are also approaches diverging from the GNN framework. \citet{zhang2022hierarchical}, for example, applied transformers to graph data using adaptive sampling, but this method did not fully address heterogeneous connection information. Our work aims to bridge this gap by developing an approach that comprehensively captures connection information and is scalable for large networks, overcoming the over-smoothing issues typical in existing methods.

%% file: LaTeX/preliminary.tex
\section{Preliminaries}

\input{LaTeX/tables/notations}

In this section, we establish formal definitions of key terminologies central to our research. For the purpose of clarity and comprehensibility, we encapsulate the mathematical notations used throughout this paper in Table~\ref{tab:notations}.

\begin{itemize}
    \item \textbf{Heterogeneous Network: }Given a list types of objects $\mathcal{V} = \{V_1, V_2,...,V_{|\mathcal{V}|}\}$, where each type $V_i$ contains $|V_i|$ nodes: $\{v_{i,1},v_{i,2},...,v_{i,|V_i|}\}$. Graph $G=\langle V,\mathcal{E}\rangle$ is defined as a 
    \textit{Heterogeneous Network} on types $\mathcal{V}$ if $V(G) = \mathcal{V}$ and $E(G) = \{v_i,v_j\}$, where $v_i,v_j \in \mathcal{V}$.

    \item \textbf{Link Prediction: }Given a subset of edges $E_s \in G$ as the training set (and potentially disconnect information as well), a link prediction model captures connection patterns that given a disjoint set of edges $E'_s, E_s \cap E'_s = \emptyset$, the model could predict $\forall e \in E'_s$, $e \in G$ or not.

    \item \textbf{Random-Walk-based Graph Sampling: }A random-walk sampled sub-graph $G_ {s} \in G$ is a sequence of movements from one node to another. 
    
    Formally, $G_{s} = \{v_1,e_{12},v_2,e_{23},v_3,...,v_{|G_{s}|}\}$ has $|G_ {s}|$ nodes and $|G_ {s}|-1$ edges, $e_{i-1,i}=\{v_{i-1},v_i\}$, where $v_{i-1},v_i \in \mathcal{V}$.

    \item \textbf{Graph Message Passing: }Conventionally, a graph neural network passes messages directly to nodes' first-order neighbors, generating representation of nodes as the parameterized sum of neighbored node representations. Formally:
         
    \begin{equation}
        h_i' = \mathcal{F}_\Theta(h_i; h_v \mbox{ for } v \in N_i).
    \end{equation}
    Here $h$ could be the node feature or the representation derived from the previous graph neural layer, $N_i$ denotes to the set of first-order neighbors of node $i$. $\mathcal{F}_\Theta()$ denotes to any graph neural operator with model parameter set $\Theta$.
    In order to propagate messages to neighbors further than first-order ones, several graph neural layers are stacked. Overmuch stacks ($> 5$) can lead to the over-smoothing issues~\citep{chen2020measuring}.


\end{itemize}

%% file: LaTeX/tables/notations.tex
\begin{table}[t]
    \centering
    \caption{Mathematical Notations}
    \begin{adjustbox}{max width=0.45\textwidth}
    \begin{tabular}{l|l}
    \hline \hline
        Symbol & Description \\
        \hline
        $\mathcal{V}$ & $\mathcal{V} = \{V_1,V_2,...,V_{|\mathcal{V}|}\}$ Set of objects  \\
        $G$ & $G=<V,E>$ A graph with nodes V and edges E \\
        $E_s,E'_s$ & $E_s,E'_s \in G$ subset of training and testing links\\
        $e_{ij}$ & $e_{ij}=<v_i,v_j> \in E$ an edge between node $i,j$\\
        $\tilde{e}_{ij}$ & $\tilde{e}_{ij}=\{e_{i,i+1},...,e_{j-1,j}\}$ Concentrated edge between node of interest\\
        $G_s$ & $G_ {s} = \{v_1,e_{12},v_2,...,v_{|G_{s}|}\}$ Trivial random-walk sample \\
        $\tilde{G}_s$ & $\tilde{G}_{s} = \{v_1, \tilde{e}_{12},v_{2},...,v_{|G'_{S}|}\}$ Concentrated random-walk sample\\
        $\mathcal{F}_\Theta()$ & General graph neural operator\\
        $[\mbox{PE}]$ & General form of position encodings\\
        $W$ & Transformation weights\\
        $N_i$ & Neighbor nodes of $i$\\
        $d_v$ & Scaling factor\\
        $k$ & Maximum number of non-interested inner nodes\\
        $L$ & Random walk length\\
        $m$ & Number of samples per head node\\
        $V_h$ & Set of head nodes\\
        $V_t$ & Set of tail nodes\\
        $X_i$ & Node feature of $i$-th node\\
        $n$ & Number of interested nodes (\textit{head} and \textit{tail} nodes)\\
        $d$ & Node feature dimension\\
        $\tilde{e}$ & Trainable connection encoding for edge type tuples\\
        $[D_i]$ & Shortest path encoding of $i$-th node\\
        $[\mbox{EDG}]$ & Position encoding for edges\\
        $\sigma()$ & Activation function\\
        $h$ & Hidden representation\\
        $I$ & Sampled sequence set\\
        $P(i)$ & Positive link set of $i$-th sequence\\
        $A(i)$ & Set of all link in $i$-th sequence\\
        $\tau$ & Temperature parameter\\
        $\alpha$ & Attention coefficient\\
        $\mathbf{a}$ & Attention weight parameter\\
        $[\ ||\ ]$ & Concatenation Operation\\
        $\mathcal{L}$ & Loss terms, including $\mathcal{L}_{obs},\mathcal{L}_{obs}$\\
        $w$ & Scaling parameter for loss functions\\
        \hline \hline
        
    \end{tabular}
    \end{adjustbox}
    \label{tab:notations}
\end{table}

%% file: LaTeX/methodology.tex
\section{Methodology}
\begin{figure*}
    \centering
     
    \includegraphics[width = 1.6\columnwidth]{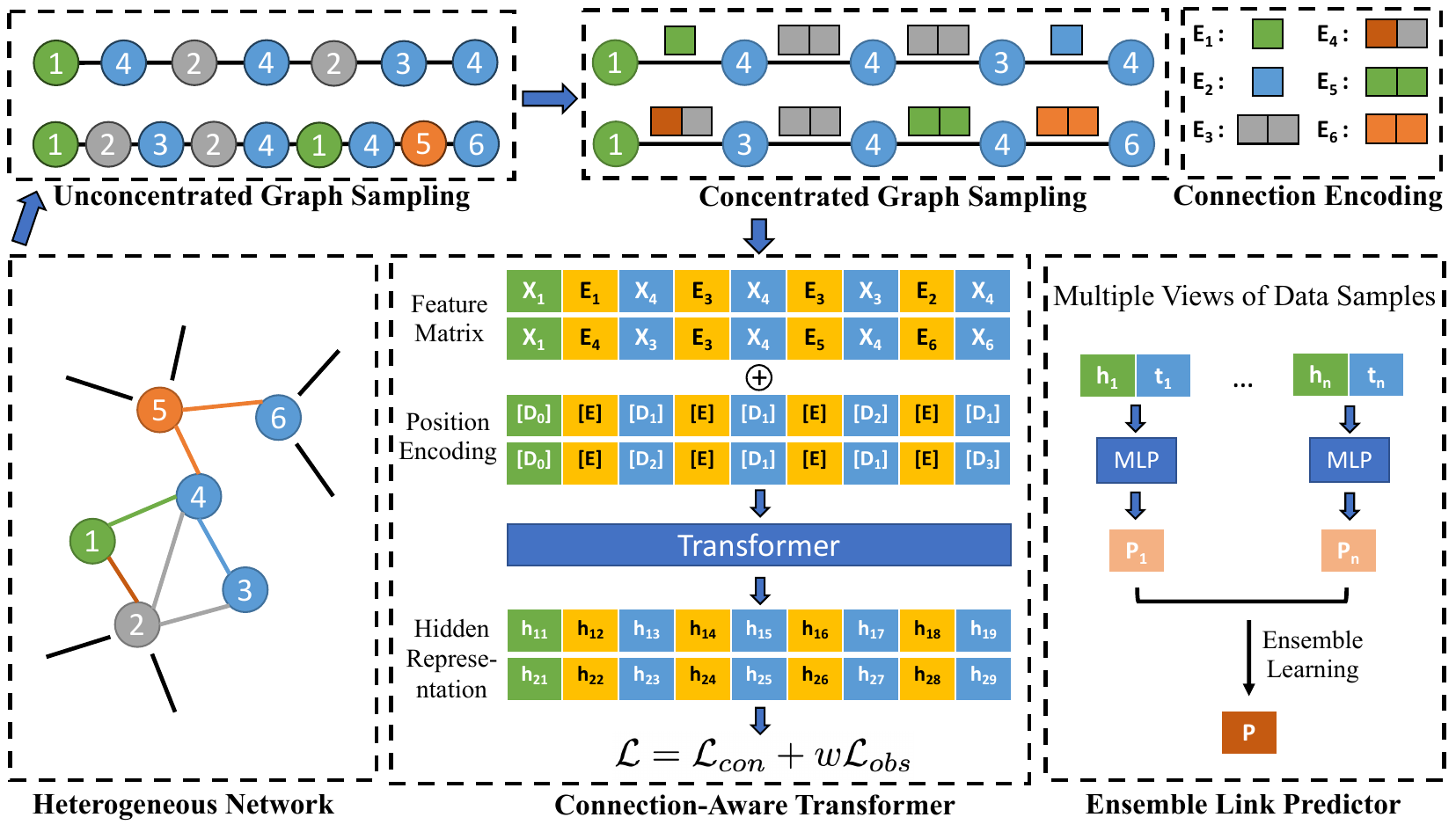}
     
    \caption{Architecture of \textbf{CHAT}. Green node $(1)$ is a \textit{head} node, blue nodes $(3,4,6)$ are \textit{tail} nodes, gray $(2)$ and orange $(5)$ node are non-interest nodes. The graph sampling technique first samples subgraph sequences from the heterogeneous network (top left), and non-interest nodes are converted into connections, i.e. tuples of edge types (top center, concatenated colored blocks). Connection encodings of the same dimension as node features are adopted w.r.t. each connections (top right), generating feature matrix and combining with position encodings as input of transformer. The connection-aware transformer is supervised by two loss functions, i.e. the contrastive link prediction loss and the observation probability loss. An ensemble link predictor is proposed for link prediction based on multiple views of data samples.}
     
    \label{fig:Architecture}
\end{figure*}

In this section, we propose a novel sampling-based graph transformer, the \underline{C}ontrastive \underline{H}eterogeneous gr\underline{A}ph \underline{T}ransformer (\textbf{CHAT}), specifically tailored for link prediction tasks within heterogeneous networks. 
The architecture of \textbf{CHAT} is shown in Figure~\ref{fig:Architecture}, which incorporates three primary components, namely concentrated graph random-walk sampling, a connection-aware transformer, and an ensemble link predictor respectively. 
We will provide details of \textbf{CHAT} in the following sections.

\subsection{Concentrated Graph Sampling}
Graph sampling is a critical component of large-scale network mining~\citep{hu2013survey}, and broadly falls into two categories: random-walk-based approaches~\citep{grover2016node2vec,dong2017metapath2vec} and k-hop neighbor-based approaches~\citep{zhang2019heterogeneous,hu2020heterogeneous}. While k-hop neighbor-based sampling strategies accentuate the relevance of closest neighbored nodes, random-walk-based techniques generate sequences of nodes, exploring both immediate neighbors and more distant nodes in the network. In the context of adapting heterogeneous networks for compatibility with transformers, we advocate for the utilization of node sequences as samples. 
If choosing neighbor-based sampling, it is ambiguous to place the sampled neighbors into sequence and the relative locations of nodes bring about permutation dependence issue. Random-walk-based sampling, on the contrary, provides an intuitive means of generating continuous walks that can naturally be treated as sampled node sequences, thus we choose to use random-walk-based sampling instead of neighbor-based ones. Formally, the sampled sequence of $i$-th node $G_{S_i}$ is denoted as:
\begin{equation}
    G_{S_i} = \{v_1,e_{12},v_2,...,v_{|G_{S_i}|}\},
\end{equation}
where $v_1$ is the $i$-th node, index $_1$ indicates node $i$ locates at the first position of the sequence. 
$e_{12} = \langle v_1, v_2 \rangle \in \mathcal{E}$ is an edge between $i$-th node ($v_1$) and its neighbor ($v_2$).

Although we have decided to use random-walk-based graph sampling technique that generate sequences of subgraphs,
directly utilizing general random walk~\citep{grover2016node2vec}
is unsuitable for the task of link predictions in heterogeneous networks. 
An alternative specialized random walk technique~\citep{dong2017metapath2vec} is designed for sampling on heterogeneous networks, yet the sampling technique only sample nodes following pre-defined meta-paths, which needs careful human selection and may easily lose crucial connection information that was not included in the pre-defined meta-paths. Defining meta-paths over complex heterogeneous network requires in-depth domain-specific knowledge as well.
To this end, we aim to develop a novel random-walk-based graph sampling technique tailored specifically for heterogeneous networks. This technique obviates the need for pre-defined meta-paths, yet adeptly captures a holistic view of connection information.

A naive approach might be to keep track of all node and edge types during sampling.
Figure~\ref{fig:Architecture} shows an example of such unconcentrated graph sampling (top left). 
A glaring drawback of this method is the unintentional inclusion of nodes and edges that aren't of primary interest (node (2),(5)), leading to an unwieldy expansion of the node space. 
However, simply discarding these seemingly irrelevant nodes is not a judicious decision either, given their integral role in furnishing pivotal connection data between the nodes we are genuinely focused on.

Along this line, we introduce a novel ``concentrated'' sampling method that effectively simplifies the non-interest nodes within heterogeneous networks
into connections.
Referring to the example shown in Figure~\ref{fig:Architecture}, node (1),(2),(3) are unique node types. An unconentrated sampling may sample a subgraph sequence starting from node (1), bypassing node (2) and visit node (3), denoted as $\{v_1, e_{12}, v_2, e_{23}, v_3\}$, where $v,e$ represents nodes and edges respectively. Under the concentrated graph sampling, only \textbf{nodes of interest} ($v_1,v_3$) are preserved, while the non-interest node $v_2$ is omitted, transforming the sequence into $\{v_1, \tilde{e}_{13}, v_3\}$, $\tilde{e}_{13} = <E(e_{12}),E(e_{23})>$, where $E(e)$ represents the edge type of $e$, and $\tilde{e}_{13}$ is the \textit{tuple of edge types} associated with $v_2$, 
treated as \textbf{connection} between $v_1$ and $v_3$. Similarly, if there are $k$ non-interest nodes between two nodes of interest, the converted connection would be a tuple of $k+1$ edge types.


Given a random-walk sampled subgraph sequence starting from $i$-th node $G_{S_i} = \{v_1,e_{12},v_2,...,v_{|G_{S_i}|}\}$, our concentrated sampling approach converts it into a simplified form $\tilde{G}_{S_i} = \{\tilde{v}_1, \tilde{e}_{12},\tilde{v}_{2},\tilde{e}_{23},...,\tilde{v}_{|\tilde{G}_{S_i}|}\}$, ensuring that every node $\tilde{v}$ in $\tilde{G}_S$ is a node of interest. 

In the context of link prediction within heterogeneous networks, the nodes of interest are those who are directly involved in the links being predicted. The originating node of a link is termed the \textit{head} node, while the terminating node is referred to as the \textit{tail} node. For example, in the realm of DTI prediction, drug nodes function as \textit{head} nodes, whereas target nodes serve as \textit{tail} nodes. We further refine our sampling protocol to ensure that in each sampled subgraph sequence, only the first node is a \textit{head} node, with all subsequent nodes being \textit{tail} nodes. This is exemplified in Figure~\ref{fig:Architecture} second-row sampling instance, where node (1) within the middle of the sequence is also treated as non-interested node. The rationale behind this is elaborated in the following section. The specifics of the concentrated graph sampling technique are detailed as pseudocode in the Appendix (Algorithm~\ref{alg:randomwalk}).

\subsection{Connection-Aware Transformer}
The proposed concentrated graph sampling technique generates a set of random-walk-based subgraph sequences, each sequence starts with a \textit{head} node, followed by a sequence of \textit{tail} nodes, connections between nodes are tuples of edge types
. Our concentrated Graph Sampling also proves to be a generalized form of meta-path-based approaches (Proof in Appendix Theorem~\ref{the:1}). In this section, we design a transformer-based model in learning comprehensive connection knowledge based on the sampled sequences.

\subsubsection{Connection-Awareness:} Traditional transformers, for example, BERT~\citep{devlin2018bert}, consider sentences as sequence of tokens, and feed the token sequence into a self-attention transformer, supervised with downstream objectives, e.g. reconstruction loss and sentence classification loss. One crucial difference between word token sequences and sampled subgraph node sequences is the connections between nodes contain unique information, while connections between word tokens make less sense, only the relative position information is useful. In consideration of the connection information, we choose to integrate both nodes and concentrated edge type tuples as input sequence of transformer. In order to do so, we need to ensure nodes and concentrated edge type tuples are of the same dimension. 

Let~$X \in \mathbb{R}^{N \times d}$ denotes to $d$-dimensional features of $N$ interested (\textit{head} and \textit{tail}) nodes. For each tupled edge type~$\tilde{e}$, we assign a trainable $d$-dimensional parameter~$[\tilde{e}]$ as its connection encoding. A random-walk subgraph sequence sampled by concentrated graph sampling~$\tilde{G}_S \in \mathbb{R}^{(2L-1) \times d}$ is now a sequence matrix with~$2L-1$ tokens and $d$ dimensional features. Formally:
\begin{equation}
    \tilde{G}_{S} = [X_1,[\tilde{e}_{12}],X_2,...,[\tilde{e}_{L-1,L}],X_L]^T.
\end{equation}
Comparing with the path formulation technique designed by~\citet{zhu2021neural} that incorporates all the inner paths within nodes, our sampling-based edge formulation avoids the scaling issue if under a dense large network setting.

Position encoding is a crucial component for transformer-based models. Conventionally a relative position-oriented encoding is utilized for textual sequences, e.g. a 2-d sinusoidal function~\citep{vaswani2017attention}. In the earlier attempts on smaller graphs, a centrality-based encoding is utilized~\citep{ying2021transformers}. However, the centrality encoding is incompatible on a large graph since a global centrality encoding could be biased on sampled subgraphs. Besides, a node centrality-oriented encoding may not contribute to the link prediction task. 
We propose to use the shortest distance between the \textit{head} node and \textit{tail} nodes as the position encodings of \textit{tail} nodes.
A zero-distance encoding is applied to the \textit{head} node itself, and a special $[\mbox{EDG}]$ encoding is applied to connections for consistency. 
A transformer takes the encoded sequence matrix as input, generating a hidden representation sequence matrix $H=[h_1,h_2,...,h_{2L-1}]^T$. Formally:
\begin{eqnarray}
    h_{2i-1} &=& \sigma(\mbox{Self-Attention}(W(X_i + [D_i]))) \\
    h_{2i} &=& \sigma(\mbox{Self-Attention}(W([\tilde{e}_{i,i+1}] + [\mbox{EDG}]))).
\end{eqnarray}
Here $[D_i]$ denotes to the shortest path encoding of $i$-th node, $\sigma()$ is an activation function, e.g. ReLU~\citep{agarap2018deep}.

\subsubsection{Objective: }Our designed objective contains two loss functions to comprehensively learn the connection (as well as disconnection) information. One is a contrastive link prediction loss, and the other is an observation probability loss.
\begin{itemize}
    \item \textbf{Contrastive Link Prediction: } Recent years have witnessed the growth of using contrastive learning scheme for improving classification robustness~\citep{khosla2020supervised,you2020graph,chen2020simple}. Link prediction is naturally a perfect fit for the task of contrastive learning, since positive links act like anchors, and negative links serve as negative samplings~\citep{zhang2023line,peng2022smile}. We follow the supervised contrastive learning objective derived from~\citet{khosla2020supervised}, and develop our contrastive link prediction loss as the following form:
    \begin{equation}
        \mathcal{L}_{con} = \sum\limits_{i \in I} \frac{-1}{|P(i)|}\sum\limits_{p \in P(i)} \mbox{log} \frac{\mbox{exp}(h_{i1} \cdot h_{ip} / \tau)}{\sum\limits_{a \in A(i)} \mbox{exp}(h_{i1} \cdot h_{ia} / \tau)}.
    \end{equation}
    Here $I$ denotes to the set of sampled sequences, $\tau$ is a scalar temperature parameter. For each sampled sequence $i$, $P(i)$ is the set of known positive links, $p \in P(i)$ indicates $p$-th \textit{tail} node has positive link with \textit{head} node in $i$-th sampled sequence. Similarly, $A(i)$ denotes to the set of all links. $h_{i1}$ means the \textit{head} node hidden representation of $i$-th sampled sequence, $h_{ia}$ denotes to hidden representation of $a$-th \textit{tail} node, respectively.

    Now we can look back to the question ``\textit{Why only \textit{tail} nodes are sampled during the concentrated graph sampling, treating visited \textit{head} nodes as inner ones?}'' left unanswered in the previous section. If preserving other \textit{head} nodes, the contrastive loss for each sampled sequence will contain links between multiple \textit{head} and \textit{tail} nodes, drawing biased training issue since the sampling distribution of observing other \textit{head} nodes cannot be guaranteed. Keeping only one \textit{head} node ensures fair training for all \textit{head} nodes.
    
    \item \textbf{Observation Probability: }The contrastive loss, though effective, considers only the relative placement of node representations without utilizing the connection information. ~\citet{grover2016node2vec} introduces an objective function that maximizes the probability of observing a network neighborhood conditioned on feature space. We improve it into an attentive connection-aware observation probability loss, strengthening the usage of connection information. Firstly, we define a pairwise connection attention $\alpha$ as:
    \begin{equation}
        \alpha_{i,i+1} = \frac{\mbox{exp}(\sigma(\mathbf{a}^T[h_{2i-1}||h_{2i}||h_{2i+1}]))}{\sum\limits_{j=1}\limits^{L-1}\mbox{exp}(\sigma(\mathbf{a}^T[h_{2j-1}||h_{2j}||h_{2j+1}]))},
        \label{eq:alpha}
    \end{equation}
    where $\mathbf{a} \in \mathbb{R}^{3d}$ is a weight vector that transfers the concatenation ($||\cdot||$) of node-connection-node tuple into scalar. Once calculated the pairwise connection attention, we can define our connection-aware observation probability loss as:
    \begin{equation}
        \mathcal{L}_{obs} = - \sum\limits_{i \in I}\sum\limits_{j=1}\limits^{L-1} \alpha_{j,j+1}(h_{2j-1} \cdot h_{2j+1}),
    \end{equation}
    where $h_{2j-1} \cdot h_{2j+1}$ is the dot product of two connected node hidden representations. Combining both loss functions, our final objective in training the connection-aware transformer is written as:
    \begin{equation}
        \mathcal{L} = \mathcal{L}_{con} + w\mathcal{L}_{obs},
    \end{equation}
    where $w$ is a scaling parameter between loss functions.
\end{itemize}

\input{LaTeX/tables/main_results}

\subsection{Ensemble Link Predictor}
A predictor aims at predicting if link exists between given queried head and tail nodes. In terms of predictor architectures, we follow the design of ``projection network'' in \citet{khosla2020supervised} that uses a multi-layer perceptron~\citep{hastie2009elements} over dot-product of head and tail node hidden representations to make predictions:
\begin{equation}
    \mbox{Predict}(v_h,v_t) = \mbox{$\sigma$ }(\sigma((h_{v_h} \cdot h_{v_t})W_1)W_2).
\end{equation}

We leave the investigation of optimal predictor architecture to future work, but keep focus on the discussion of prediction scheme. As mentioned by ~\citet{hinton2015distilling}, ``A very simple way to improve the performance of almost any machine learning algorithm is to train many different models on the same data and then to average their predictions.'' We consider the other side of this sentence, i.e. make predictions based on multiple instances of data samples and then to average the predictions, which is also a common procedure in terms of ensemble learning~\citep{sagi2018ensemble,dong2020survey}. Formally, the ensembled prediction is:
\begin{equation}
    \mbox{Ensemble}(v_h,v_t) = \frac{1}{Z}\sum\ _{v_t \in A(v_h)} \mbox{ Predict}(v_h,v_t).
\end{equation}
Here $A(v_h)$ denotes to the sampled sequences starting at~$v_h$, $Z$ represents the total times of observing $v_t$ in $A(v_h)$. If the frequency of observing $v_t$ in the sampled sequence of $v_h$ is zero, the average of hidden representations is used instead of averaging predictions. 

%% file: LaTeX/tables/main_results.tex
\begin{table*}[t]
    \centering
    \footnotesize
         
    \caption{The overall performance of different models on three datasets.}
         
    \begin{adjustbox}{max width=1\textwidth}
    \begin{tabular}[t]{c|cccc|cccc|cccc}\toprule
    \textbf{Dataset} & \multicolumn{4}{c|}{DTI-315} & \multicolumn{4}{c|}{DTI-708} & \multicolumn{4}{c}{DTI-258K} \\ \midrule
    \multirow{2}*{\textbf{Algorithm}} & \multicolumn{2}{c|}{\textbf{Classification Metrics }} & \multicolumn{2}{c|}{\textbf{Ranking Metrics}}  & \multicolumn{2}{c|}{\textbf{Classification Metrics }} & \multicolumn{2}{c|}{\textbf{Ranking Metrics}}  & \multicolumn{2}{c|}{\textbf{Classification Metrics }} & \multicolumn{2}{c}{\textbf{Ranking Metrics}}  \\
    \cmidrule{2-13}
     & $Accuracy$ & $F$-$1$ & $AUC$ & $AUPR$ & $Accuracy$ & $F$-$1$ & $AUC$ & $AUPR$ & $Accuracy$ & $F$-$1$ & $AUC$ & $AUPR$ \\ \midrule 
    
    Random Forest & 98.36\% & 16.86\% & 86.76\% & 29.20\% & 74.79\% & 66.33\% & 92.83\% & 94.32\% & 86.18\% & 71.23\% & 92.94\% & 87.67\% \\
Logistic Regression & 98.51\% & 38.32\% & 90.90\% & 42.42\% & 85.39\% & 83.15\% & 92.95\% & 94.90\% & 84.21\% & 64.71\% & 86.25\% & 82.32\% \\
SMPSL & 98.07\% & 41.98\% & 92.96\% & 39.91\% & 71.95\% & 71.95\% & 80.74\% & 81.15\% & 70.39\% & 41.56\% & 63.47\% & 45.57\% \\
DTINet & 48.47\% & 5.23\% & 80.46\% & 38.45\% & 68.70\% & 75.23\% & 89.92\% & 92.08\% & 47.37\% & 38.46\% & 46.29\% & 27.58\% \\
\midrule
Metapath2Vec & 97.36\% & 19.38\% & 86.08\% & 13.31\% & 85.94\% & 85.94\% & 92.69\% & 93.85\% & 75.66\% & 47.59\% & 62.40\% & 54.83\% \\
HAN & 97.16\% & 46.15\% & 93.60\% & 49.89\% & 84.40\% & 85.19\% & 87.63\% & 78.82\% & 88.16\% & 75.68\% & 93.02\% & 87.95\% \\
ANS-GT & 97.89\% & 53.37\% & 94.07\% & 62.72\% & 89.73\% & 89.86\% & 93.73\% & 92.96\% & 95.32\% & 91.46\% & 96.18\% & 90.92\% \\
\midrule
EEG-DTI & 96.63\% & 44.44\% & 95.30\% & 60.24\% & 92.68\% & 92.13\% & 95.26\% & 96.34\% & 93.47\% & 88.95\% & 94.02\% & 88.10\% \\
SGCL-DTI & 95.64\% & 38.64\% & 94.89\% & 52.84\% & 91.69\% & 91.71\% & 95.26\% & 95.80\% & 94.19\% & 90.43\% & 95.74\% & 89.53\% \\
MHGNN-DTI & 96.05\% & 41.43\% & 95.79\% & 62.99\% & 92.71\% & 92.79\% & \textbf{96.93\%} & 95.52\% & 94.84\% & 90.97\% & 95.68\% & 90.79\% \\
\midrule
\textbf{CHAT } & \textbf{98.56\%} & \textbf{53.95\%} & \textbf{96.22\%} & \textbf{64.77\%} & \textbf{93.49\%} & \textbf{93.53\%} & 96.87\% & \textbf{96.75\%} & \textbf{95.63\%} & \textbf{92.11\%} & \textbf{96.92\%} & \textbf{91.32\%}\\
    
    \bottomrule
    \end{tabular}
    \end{adjustbox}
    \label{tab:overallperf1}
\end{table*}

%% file: LaTeX/Experiments.tex
\section{Experiment}

\input{LaTeX/Dataset}

\subsection{Experimental Results}

\subsubsection{Overall Performance }
Table~\ref{tab:overallperf1} presents a comparative analysis of our \textbf{CHAT} model versus the established baselines across three datasets, evaluated using four distinct metrics. Optimal values in each column are highlighted in bold. It is evident that the \textbf{CHAT} model exhibits superior performance, surpassing the baseline models in the majority of the evaluations. Notable observations gleaned from the table include: 
(i) Approaches that leverage domain-specific knowledge, such as \textbf{EEG-DTI}, \textbf{SGCL-DTI}, and \textbf{MHGNN-DTI}, exhibit superior performance in the DTI-708 dataset when compared to generic methodologies. This underscores the potency of domain-informed strategies. Remarkably, despite being devoid of domain-specific insights, our proposed \textbf{CHAT} model surpasses even these domain-centric methods, thereby attesting to the efficacy and adaptability of our approach.
(ii) While Logistic Regression yields the highest accuracy on the DTI-315 dataset, its performance on the other three metrics remains suboptimal. This suggests that Logistic Regression may primarily capture the overarching label distribution without effectively classifying or ranking individual links with precision. Logistic Regression's performance on the other two datasets further indicates its inability to excel in more balanced situations. (iii) Graph neural network-centric strategies, in conjunction with transformer-based methodologies, significantly surpass non-deep-learning techniques. This underscores the potency of deep learning models in harnessing graphical data. We have also tested our approach over 
 link prediction tasks on three different domains. The results and analysis are included in the Appendix (Table~\ref{tab:all-domain}).

 \subsection{Additional Experiments}

\input{LaTeX/tables/all-domain}

In this subsection, we present extended experimental results comparing our CHAT model with various baseline methods across multiple domains. We employed three diverse heterogeneous network datasets for this analysis: ACM, DBLP, and IMDB, which are publicly available for reference and use\footnote{\url{https://github.com/Jhy1993/HAN/tree/master/data}}. Our evaluation metrics include the AUC and AUPR scores for each dataset. As shown in Table~\ref{tab:all-domain}, the CHAT model demonstrates a significant improvement over all baseline methods. This enhancement is not only evident when compared to the results in Table~\ref{tab:overallperf1}, where CHAT surpasses DTI prediction-specialized baselines, but it is also more pronounced in Table~\ref{tab:all-domain}. The superior performance in these cases can be attributed to CHAT's advanced capability in assimilating domain-specific knowledge, setting it apart from other general-purpose approaches.

\subsubsection{Ablation Study}
\begin{figure}[t]
    \centering
        
    \includegraphics[width = 1.0\columnwidth]{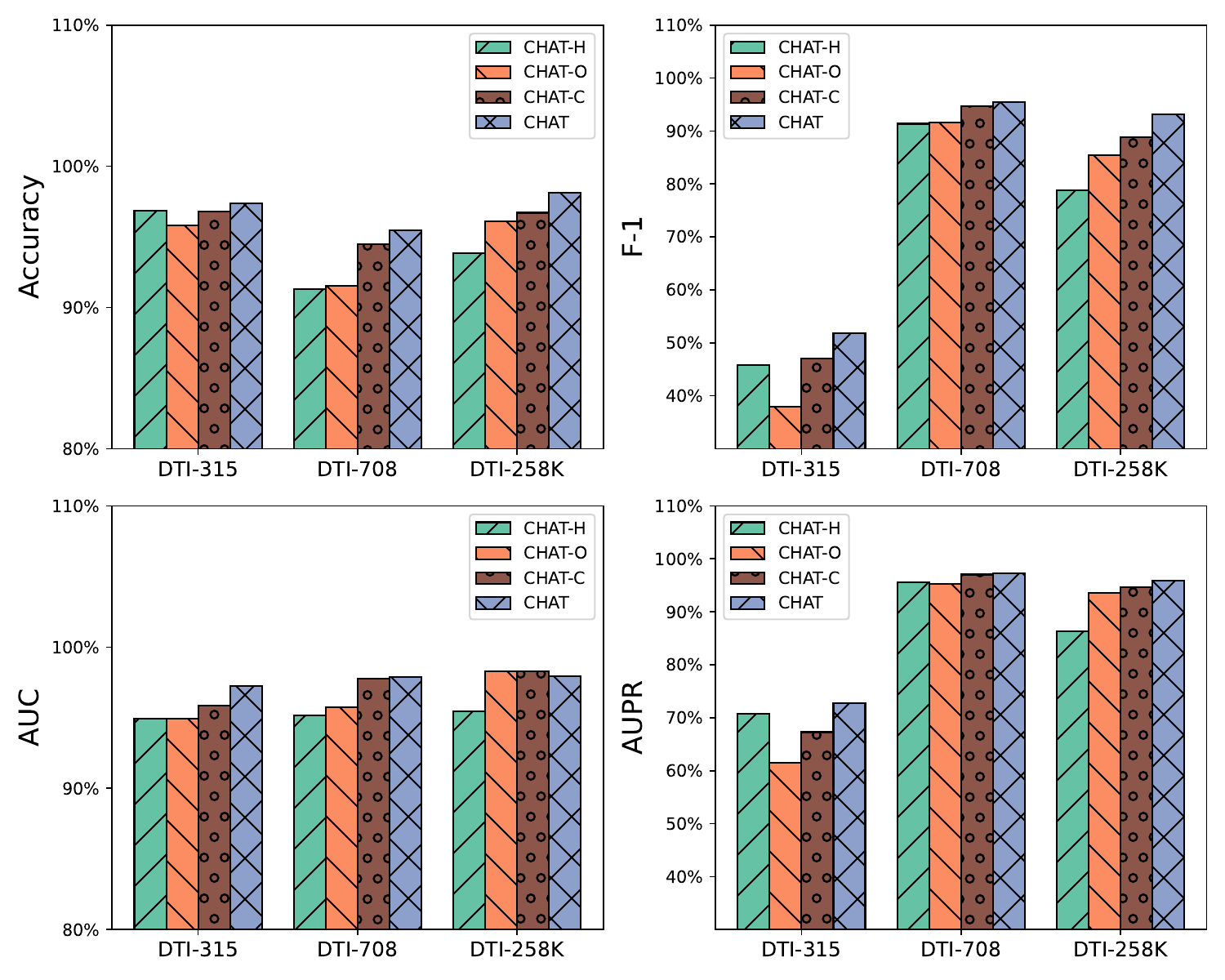}
        
    \caption{Ablation studies on three datasets.}
        
    \label{fig:ablation}
\end{figure}

To scrutinize the contribution of individual modules within the \textbf{CHAT} framework, we conduct an ablation study focusing on two critical modules: the heterogeneous connection module and the contrastive learning module. Specifically, \textbf{CHAT-H} represents the \textbf{CHAT} model with the exclusion of the heterogeneous connection module—this is achieved by maintaining uniform connections between nodes. \textbf{CHAT-O} represents the \textbf{CHAT} model without the observation loss; Conversely, \textbf{CHAT-C} represents the \textbf{CHAT} model without the contrastive loss.
Figure~\ref{fig:ablation} presents a comparative assessment of the \textbf{CHAT} against its ablated versions across four evaluation metrics for all three datasets. A discernible performance drop is evident upon excluding the heterogeneous connection module, as seen in the contrast between green and blue bars. Similar pattern can be observed by ablating the observation loss. Furthermore, the removal of the contrastive loss function is particularly impactful in the context of imbalanced labels, as observed in the distinction between orange and blue bars for the DTI-315 dataset. Overall, the comprehensive \textbf{CHAT} framework leverages the synergistic benefits of both modules to achieve superior performance.

\subsubsection{Interpretability Study}
\begin{figure}[t]
    \centering
        
    \includegraphics[width = 1\columnwidth]{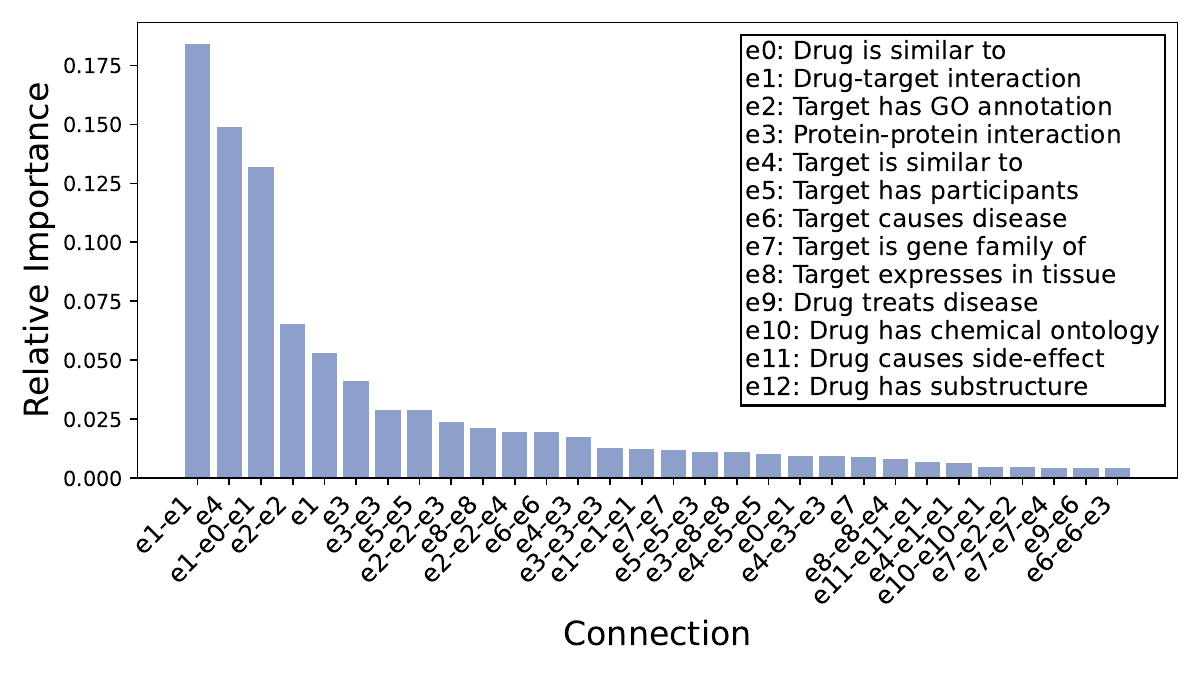}
        
    \caption{Relative importance of top-30 connections.}
    \label{fig:interpretability}
\end{figure}

One of the distinguishing features of our proposed \textbf{CHAT} model lies in its capacity to seamlessly integrate diverse connections between nodes of interest, circumventing the necessity for pre-specified meta-paths. The relative significance of these connections can be ascertained using a softmax function over all projected connection representations, as delineated in Equation~\ref{eq:alpha}. The visualization in Figure~\ref{fig:interpretability} elucidates the relative importance of the top 30 connections on dataset DTI-258K. Each of these connections, intriguingly, comprises no more than three edges and establishes links between either drug-target nodes or target-target nodes. To illustrate, the highest ranked connection, denoted as ``e1-e1'', symbolizes a target-drug-target connection defined by drug-target interactions, while the connection labeled ``e9-e6'' epitomizes a drug-disease-target connection. A close inspection reveals that the similarity metrics (e0, e4), interaction metrics (e1, e3), and GO annotation metric (e4) dominate the landscape, featuring in 8 of the top 10 connections.

\subsubsection{Sensitivity Analysis}

\begin{figure}[t]
    \centering
        
    \hspace*{0\columnwidth}
    \includegraphics[width = 0.75\columnwidth]{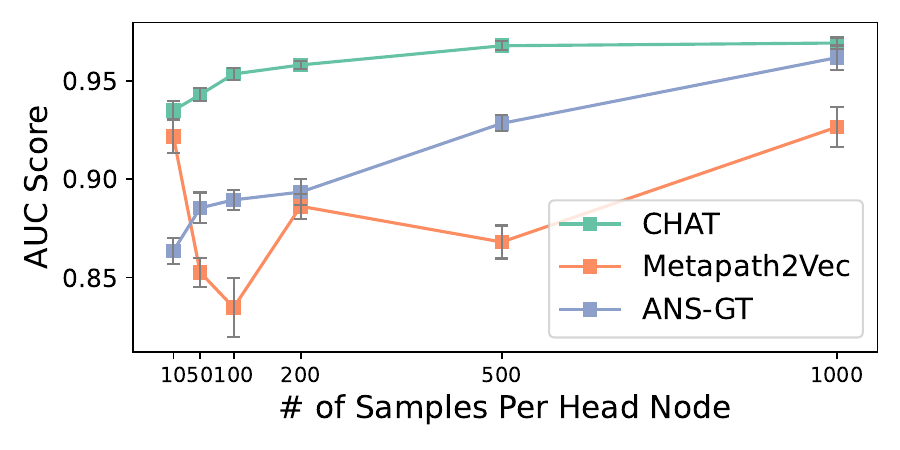}
        
    \caption{Sensitivity analysis of sample size.}
        
    \label{fig:sensitivity}
\end{figure}

Within the context of sampling-based methodologies, the computational overhead associated with \textbf{CHAT} is contingent upon the sampling size designated per head node. As delineated in Figure~\ref{fig:sensitivity}, the variation in AUC scores, consequent to adjustments in the number of samples per head node on the DTI-258K dataset, is showcased for three distinct sampling-based models: \textbf{CHAT}, \textbf{Metapath2Vec}, and \textbf{ANS-GT}. A discernible trend indicates performance augmentation and stabilization for \textbf{CHAT} and \textbf{ANS-GT} as the sample size escalates, albeit \textbf{Metapath2Vec} manifests sporadic fluctuations. Intriguingly, \textbf{CHAT} achieves convergence at an earlier juncture (around the 100 samples mark) compared to \textbf{ANS-GT}. Furthermore, \textbf{CHAT} exhibits relatively smaller variance, as indicated by the narrower vertical bars. 

\subsection{Scalability Analysis}

We provide a theoretical scalability analysis for our proposed CHAT model.
In terms of computational complexity, CHAT derives from the original transformer model. Let~$n$ be the sample node length under general random-walk-based sampling, the time complexity of self-attention module is $\mathcal{O}(n^2d + nd^2)$, where~$d$ is the embedding dimension.

Our concentrated sampling technique adopts concentrated sampling of nodes only of interest, with maximum $k$ inner nodes. In the worst-case, all nodes are of interest and our concentrated sampling technique devolves to non-concentrated sampling. If under an average scenario, let~$\mathbb{E}_k$ denotes to the expected number of inner nodes under maximum inner node tolerance~$k$, our proposed concentrated graph sampling only need sample ~$n/\mathbb{E}_k$ nodes to achieve the same number of sampled nodes of interest comparing to non-concentrated sampling, thus alleviating the time complexity to~$\mathcal{O}((n/\mathbb{E}_k)^2d + (n/\mathbb{E}_k)d^2)$.

In terms of our connection-aware transformer module, since it incorporates additional connection encodings, with one time additional connection tokens, thus the time complexity of CHAT's self-attention module is updated as $\mathcal{O}((2n/\mathbb{E}_k)^2d + (2n/\mathbb{E}_k)d^2)$.

Comparing with sampling-based graph transformers versus non-sampling graph transformer approaches, A sampling-based graph transformer approach needs conducting self-attentions over all sampled node sequences, while non-sampling approaches need substitute the sequence length from sample sequence length~$n$ to node space size~$N$. Specifically, the sampling-based graph transformer approach multiplies the sample size $S$, with the total time complexity of finishing one epoch of training of our proposed CHAT updated to $\mathcal{O}(S(2n/\mathbb{E}_k)^2d + (2n/\mathbb{E}_k)d^2)$. For non-sampling approach without connection-awareness, the time complexity is updated to $\mathcal{O}(N^2d + Nd^2)$.

From the comparison of both time complexities, it is evident that non-sampling approach is more effective under smaller node space (smaller~$N$), while it is infeasible for non-sampling approach over large-scale network with both time and memory-usage complexity concerns.


\section{Conclusion}
We focused on the challenge of predicting latent links within heterogeneous networks. Addressing two key limitations with existing link prediction methods—specifically, the over-smoothing issue associated with GNN-based models and the requirement of manually defining meta-paths for heterogeneous network approaches—we introduced the \underline{C}ontrastive \underline{H}eterogeneous gr\underline{A}ph \underline{T}ransformer (\textbf{CHAT}).
We proposed a concentrated graph sampling technique designed to explore all potential connections, eliminating the need for human-defined meta-paths. Furthermore, we incorporated a connection-aware transformer that was specifically designed to integrate heterogeneous connections within the transformer architecture, while concurrently mitigating the over-smoothing concerns. To augment this, we introduced a dual-faceted loss function, alongside an ensemble link predictor, to collectively guide the connection-aware transformer in its operations.
Our rigorous experiments are conducted on three drug-target interaction prediction datasets, benchmarked against ten distinct baseline methods, provided a deep insight into the effectiveness of \textbf{CHAT}. 

\section{Ethics Statement}

This research was conducted with an unwavering commitment to ethical standards, not only in methodology but also in considering the broader impact of our work. We ensured data integrity, transparency, and compliance with all relevant ethical and legal regulations. No direct human or animal subjects were involved, and we adhered to the no harm principle, mindful of our research's potential influence on future medical and pharmacological applications. All data were ethically sourced, and confidentiality and privacy were stringently maintained. This manuscript is original, has not been published before, and is not under consideration elsewhere. We have adhered to the highest standards in scientific publishing. This study, while technical in nature, aspires to contribute meaningfully to the advancement of drug-target interaction understanding and to have a positive, far-reaching impact on public health and medical research, fostering ethical applications of broader network analysis.

%% file: LaTeX/Dataset.tex
\subsection{Dataset Description}

Our empirical investigation begins with the assembly of real-world link prediction tasks across heterogeneous networks. Among such tasks, drug-target interaction prediction on biological networks stands as one of the most researched, offering an abundance of baseline comparisons and publicly available datasets. In consideration of several potential candidates, we have opted to work with three publicly accessible and widely recognized datasets that have undergone rigorous peer scrutiny. 
\textbf{We have opted to focus on a single type of link prediction task to evaluate whether CHAT can surpass approaches that leverage domain-specific knowledge}. This decision is driven by one of CHAT's key technical claims: its ability to function effectively without relying on domain-specific expertise. We also conduct experiments over different domains to test the effectiveness of \textbf{CHAT}. Please refer to Table~\ref{tab:all-domain} for more details.

Our selected three publicly available datasets are DTI-315~\citep{perlman2011combining}, DTI-708~\citep{luo2017network} and DTI-258K~\citep{fu2016predicting}. The statistics of the chosen datasets are outlined in Table~\ref{tab:dataset}.
\input{LaTeX/tables/statistics}

\subsection{Experimental Settings}
\textbf{Baselines: }In order to provide a comprehensive evaluation of our proposed \textbf{CHAT} model, we contrast its performance with a diverse array of baseline algorithms, encompassing both conventional methodologies and state-of-the-art approaches. These baselines can be partitioned into four distinct categories: (i) Techniques that leverage meta-path counts as features (\textbf{Meta-path+Logistic Regression}, \textbf{Meta-path+Random Forest}~\citep{fu2016predicting}, and \textbf{SMPSL}~\citep{zhang2023meta}). (ii) Method that employ matrix factorization and projection techniques (\textbf{DTINet}~\citep{luo2017network}). (iii) Generalized deep learning approaches to link prediction (\textbf{Metapath2vec}~\citep{dong2017metapath2vec}, \textbf{HAN}~\citep{wang2019heterogeneous}, and \textbf{ANS-GT}~\citep{zhang2022hierarchical}). (iv) Deep learning-based approaches explicitly designed for drug-target interaction (DTI) prediction (\textbf{EEG-DTI}~\citep{peng2021end}, \textbf{MHGNN}~\citep{li2023metapath}, and \textbf{SGCL-DTI}~\citep{li2022supervised}). 

\textbf{Evaluation Metrics: } We conduct experiments over two categories of evaluation metrics: (i) The classification metrics, including Accuracy score and F-1 score; (ii) The ranking metrics, including Area under ROC curve score (AUC) and Area under precision-recall curve score (AUPR). All  metrics are commonly used by related scholars. Details concerning the datasets, baselines, evaluation protocols, and the experimental setups are available in the Appendix. The source code of CHAT is publically available at: \href{https://anonymous.4open.science/r/CHAT-8978/}{anonymous.4open.science/r/CHAT-8978}.

%% file: LaTeX/tables/statistics.tex
\begin{table}[h]
    \centering
    \caption{Statistics of three datasets.}
    \begin{adjustbox}{max width=0.4\textwidth}
    \begin{tabular}{l|c|c|c}
    \toprule
        &DTI-315 & DTI-708 & DTI-258K\\
        \midrule
        \# Node Types & 2 & 4 & 9\\
        \# Edge Types & 9 & 7 & 11\\
        \# Drugs & 315 & 708 & 258,030\\
        \# Targets & 250 & 1,512 & 22,056\\
        \# Known Interactions & 1,306 & 1,923 & 188,781\\
        \bottomrule
        
    \end{tabular}
    \end{adjustbox}
    
    \label{tab:dataset}
\end{table}

%% file: LaTeX/tables/all-domain.tex
\begin{table}[h]
    \centering
    \caption{AUC and AUPR Scores over various domain datasets}
    \begin{adjustbox}{max width=0.45\textwidth}
    \begin{tabular}{c|cc|cc|cc}
    \toprule
        \textbf{Dataset}&\multicolumn{2}{|c}{ACM}& \multicolumn{2}{|c}{DBLP} & \multicolumn{2}{|c}{IMDB}\\
        \midrule
 \textbf{Algorithm}& $AUC$& $AUPR$& $AUC$& $AUPR$& $AUC$& $AUPR$\\
        \midrule
        RFs& 82.31\%& 82.01\%& 82.73\%& 82.99\%&  66.99\%& 59.54\%\\
        LR& 83.42\%& 83.73\%& 85.30\%& 85.53\%&  62.96\%& 57.66\%\\
        SMPSL& 80.36\%& 80.08\%& 75.65\%& 76.23\%&  59.98\%& 56.93\%\\
 MP2Vec& 85.59\%& 86.33\%& 87.71\%& 88.35\%& 65.66\%& 60.88\%\\
 HAN& 88.06\%& 89.07\%& 90.36\%& 91.06\%& 71.47\%& 68.49\%\\
 ANS-GT& 88.37\%& 90.05\%& 92.85\%& 94.00\%& 75.15\%& 71.29\%\\
 \midrule
 CHAT& \textbf{91.31}\%& \textbf{92.26}\%& \textbf{94.27}\%& \textbf{95.22}\%& \textbf{79.41}\%& \textbf{76.75}\%\\
 \bottomrule
        
    \end{tabular}
    \end{adjustbox}
    
    \label{tab:all-domain}
\end{table}

%% file: LaTeX/appendix.tex
\appendix
\section{Appendix}

\input{LaTeX/algorithms/randomwalk}




\begin{theorem}
The Concentrated Graph Sampling is a generalized form of meta-path-based approaches.
\label{the:1}
\end{theorem}

A meta-path-based approach captures node proximities w.r.t. pre-defined semantic meta-paths. For example, if there is a meta-path $M = \{V_1 \xrightarrow{R_1} V_2 \xrightarrow{R_2} ... \xrightarrow{R_{L-1}} V_{L}\}$ connecting node type $V_1$ and $V_L$, where $R$ denotes to edge type, it indicates there is a closeness information between interested node type $V_1$ and $V_L$ following $M$. 

Let $I$ denotes to the index of $i$-th interested node type over $M$, $|I| = n$, $V_{I_1} = V_1$ and $V_{I_n} = V_{L}$. The original metapath $M$ can be rewrite as a consecutive union of sub-meta-paths:
\begin{equation}
    M = \bigcup_{i=1}^{n}\ \{V_{I_i} \xrightarrow{R_{I_i}}V_{(I_i+1)} \xrightarrow{R_{(I_i+1)}},...,V_{I_{(i+1)}}\},
\end{equation}
where each sub-meta-path starts and ends with node of interests, all the inner nodes are out of interest. The maximum inner connection length is $sup\{I_{(i+1)} - I_i; \forall i=1,2,...,n\}$. We can see that the sub-meta-path is equivalent to a sub-sequence extracted by our proposed Concentrated Graph Sampling, if the maximum inner connection length $k$ ensures $k \geq sup\{I_{(i+1)} - I_i; \forall i=1,2,...,n\}$. Similarly, a meta-path consisting of $n$ sub-meta-paths is equivalent to the length-$n$ preamble sub-sequence extracted by Concentrated Graph Sampling as long as the random walk length $L$ ensures $L \geq n$. To this end, we successfully prove that meta-path-based approaches can be generalized to our proposed Concentrated Graph Sampling.

\subsection{Detailed Experiment Settings}

For Meta-path-based approaches, we follow the list of meta-paths of extant works. Specifically, for DTI-315, we follow~\citet{zhang2023meta} that defines five drug similarity meta-paths and three target similarity meta-paths. For DTI-708, we follow~\citet{li2023metapath,li2022supervised} that defines a total of nine meta-paths. For DTI-258K, we follow~\citet{fu2016predicting} that takes 51 meta-paths together with 51 shortest path metrics, in total 102 dimensional features as the input for meta-path-as-feature-based approaches (\textbf{Meta-path+Logistic Regression}, \textbf{Meta-path+Random Forest}, and \textbf{SMPSL}). For deep-learning-based approaches, we follow the patterns in other two datasets that explores meta-paths w.r.t. similarities and additional node types, generating a total of 9 meta-paths. Baselines that utilize pre-defined meta-paths include \textbf{Meta-path+Logistic Regression}, \textbf{Meta-path+Random Forest},  \textbf{SMPSL}, \textbf{Metapath2Vec}, \textbf{HAN}, \textbf{SGCL-DTI}, \textbf{MHGNN-DTI}.

For sampling-based approaches (\textbf{Metapath2Vec},\textbf{ANS-GT},\textbf{CHAT}, we sample 1000 sequences per head node using each method's corresponding sampling technique. For \textbf{Metapath2Vec}, the repeat path times is set to 100, and for \textbf{CHAT}, the maximum explored sequence length is also set to 100.

In terms of dataset setting, for DTI-315, a 10-fold cross-validation is conducted (consistent to \citet{zhang2023meta}, for DTI-708, a five-fold cross-validation is conducted (consistent to~\citet{li2023metapath,li2022supervised,luo2017network}, and for DTI-258K, 10 runs are conducted under a pre-splitted train-test sets (provided by~\citet{fu2016predicting}). The positive-negative link ratio for each dataset is approximately 1:29 (DTI-315), 1:1 (DTI-708) and 1:5 (DTI-258K). For dataset DTI-258K, a sampling of only nodes under interests are conducted for GNN-based approaches due to scalability, while evaluation metrics are calculated under a fair setting to other approaches. For the scaling parameter of observation probability loss $w$, we tested it from $0.1$ to $100$ during our experiments, and we report results when $w=1$ for all datasets to avoid over-tuning the scaling parameter.

For all deep learning-based baselines, the hyperparameters are tuned to the best of our attempts, meanwhile ensures a fair comparison. All the initial embeddings are randomly initialized, with 512 dimensions. All hidden state embedding sizes are set to 256 and the output representations are set to 128. For transformer-based approach, the number of layers is set to 4, and the number of heads is set to 8. The maximum training epochs is set to 1000, with each method's corresponding early-stopping triggers (if applicable).

Experiments are conducted using Python 3.10 with PyTorch. All baseline approaches are based on public version if available. We conduct experiments on a CentOS server with Intel(R) Xeon(R) Gold 6148 CPUs @ 2.40GHz and a Tesla V100 GPU with 526 GB memory. The maximum memory usage of CHAT is less than 8 GB.


%% file: LaTeX/algorithms/randomwalk.tex
\begin{algorithm}
    \small
    \caption{Concentrated Graph Sampling}
    \begin{algorithmic}\label{alg:randomwalk}
        \REQUIRE $G,k,L,m,V_h,V_t$
        \STATE \# G the original graph,
        \STATE \# k the maximum inner connections,
        \STATE \# L the random walk length,
        \STATE \# m number of samples per head node,
        \STATE \# $V_h$ set of head nodes,
        \STATE \# $V_t$ set of tail nodes.
        \STATE samples $ \leftarrow \emptyset$
        \FOR{$v_h \in V_h$}
            \STATE sample $\leftarrow \{v_h\}$
            \WHILE{sample length not reaching $L$}
                \STATE $x \leftarrow $ last node of sample
                \STATE $\tilde{e}_x,y \leftarrow $ sample a tail node $y$ using concentrated sampling starting from $x$, $\tilde{e}_x$ is the sampled edge type tuple
                \STATE $\tilde{e}_x,y \leftarrow $ resample if inner connections exceed $k$
                \STATE $\tilde{e}_x,y \leftarrow \emptyset$ if no valid samples
                \STATE sample $ \leftarrow$ sample $ \cup \{\tilde{e}_x,y\}$
            \ENDWHILE
            \STATE samples $ \leftarrow$ samples $ \cup \{\mbox{sample}\}$
            \STATE Finish if number of samples reach $m$
        \ENDFOR
        \RETURN samples
    \end{algorithmic}
\end{algorithm}

%% file: A_main.bbl

\begin{thebibliography}{51}


\ifx \showCODEN    \undefined \def \showCODEN     #1{\unskip}     \fi
\ifx \showDOI      \undefined \def \showDOI       #1{#1}\fi
\ifx \showISBNx    \undefined \def \showISBNx     #1{\unskip}     \fi
\ifx \showISBNxiii \undefined \def \showISBNxiii  #1{\unskip}     \fi
\ifx \showISSN     \undefined \def \showISSN      #1{\unskip}     \fi
\ifx \showLCCN     \undefined \def \showLCCN      #1{\unskip}     \fi
\ifx \shownote     \undefined \def \shownote      #1{#1}          \fi
\ifx \showarticletitle \undefined \def \showarticletitle #1{#1}   \fi
\ifx \showURL      \undefined \def \showURL       {\relax}        \fi
\providecommand\bibfield[2]{#2}
\providecommand\bibinfo[2]{#2}
\providecommand\natexlab[1]{#1}
\providecommand\showeprint[2][]{arXiv:#2}

\bibitem[Agarap(2018)]%
        {agarap2018deep}
\bibfield{author}{\bibinfo{person}{Abien~Fred Agarap}.}
  \bibinfo{year}{2018}\natexlab{}.
\newblock \showarticletitle{Deep learning using rectified linear units (relu)}.
\newblock \bibinfo{journal}{\emph{arXiv preprint arXiv:1803.08375}}
  (\bibinfo{year}{2018}).
\newblock


\bibitem[Bach et~al\mbox{.}(2017)]%
        {bach2017hinge}
\bibfield{author}{\bibinfo{person}{Stephen~H Bach}, \bibinfo{person}{Matthias
  Broecheler}, \bibinfo{person}{Bert Huang}, {and} \bibinfo{person}{Lise
  Getoor}.} \bibinfo{year}{2017}\natexlab{}.
\newblock \showarticletitle{Hinge-loss markov random fields and probabilistic
  soft logic}.
\newblock  (\bibinfo{year}{2017}).
\newblock


\bibitem[Borgatti et~al\mbox{.}(2009)]%
        {borgatti2009network}
\bibfield{author}{\bibinfo{person}{Stephen~P Borgatti}, \bibinfo{person}{Ajay
  Mehra}, \bibinfo{person}{Daniel~J Brass}, {and} \bibinfo{person}{Giuseppe
  Labianca}.} \bibinfo{year}{2009}\natexlab{}.
\newblock \showarticletitle{Network analysis in the social sciences}.
\newblock \bibinfo{journal}{\emph{science}} \bibinfo{volume}{323},
  \bibinfo{number}{5916} (\bibinfo{year}{2009}), \bibinfo{pages}{892--895}.
\newblock


\bibitem[Cai and Ji(2020)]%
        {cai2020multi}
\bibfield{author}{\bibinfo{person}{Lei Cai} {and} \bibinfo{person}{Shuiwang
  Ji}.} \bibinfo{year}{2020}\natexlab{}.
\newblock \showarticletitle{A multi-scale approach for graph link prediction}.
  In \bibinfo{booktitle}{\emph{Proceedings of the AAAI conference on artificial
  intelligence}}, Vol.~\bibinfo{volume}{34}. \bibinfo{pages}{3308--3315}.
\newblock


\bibitem[Cai et~al\mbox{.}(2021)]%
        {cai2021line}
\bibfield{author}{\bibinfo{person}{Lei Cai}, \bibinfo{person}{Jundong Li},
  \bibinfo{person}{Jie Wang}, {and} \bibinfo{person}{Shuiwang Ji}.}
  \bibinfo{year}{2021}\natexlab{}.
\newblock \showarticletitle{Line graph neural networks for link prediction}.
\newblock \bibinfo{journal}{\emph{IEEE Transactions on Pattern Analysis and
  Machine Intelligence}} \bibinfo{volume}{44}, \bibinfo{number}{9}
  (\bibinfo{year}{2021}), \bibinfo{pages}{5103--5113}.
\newblock


\bibitem[Chen et~al\mbox{.}(2020b)]%
        {chen2020measuring}
\bibfield{author}{\bibinfo{person}{Deli Chen}, \bibinfo{person}{Yankai Lin},
  \bibinfo{person}{Wei Li}, \bibinfo{person}{Peng Li}, \bibinfo{person}{Jie
  Zhou}, {and} \bibinfo{person}{Xu Sun}.} \bibinfo{year}{2020}\natexlab{b}.
\newblock \showarticletitle{Measuring and relieving the over-smoothing problem
  for graph neural networks from the topological view}. In
  \bibinfo{booktitle}{\emph{Proceedings of the AAAI conference on artificial
  intelligence}}, Vol.~\bibinfo{volume}{34}. \bibinfo{pages}{3438--3445}.
\newblock


\bibitem[Chen et~al\mbox{.}(2020a)]%
        {chen2020simple}
\bibfield{author}{\bibinfo{person}{Ting Chen}, \bibinfo{person}{Simon
  Kornblith}, \bibinfo{person}{Mohammad Norouzi}, {and}
  \bibinfo{person}{Geoffrey Hinton}.} \bibinfo{year}{2020}\natexlab{a}.
\newblock \showarticletitle{A simple framework for contrastive learning of
  visual representations}. In \bibinfo{booktitle}{\emph{International
  conference on machine learning}}. PMLR, \bibinfo{pages}{1597--1607}.
\newblock


\bibitem[Devlin et~al\mbox{.}(2018)]%
        {devlin2018bert}
\bibfield{author}{\bibinfo{person}{Jacob Devlin}, \bibinfo{person}{Ming-Wei
  Chang}, \bibinfo{person}{Kenton Lee}, {and} \bibinfo{person}{Kristina
  Toutanova}.} \bibinfo{year}{2018}\natexlab{}.
\newblock \showarticletitle{Bert: Pre-training of deep bidirectional
  transformers for language understanding}.
\newblock \bibinfo{journal}{\emph{arXiv preprint arXiv:1810.04805}}
  (\bibinfo{year}{2018}).
\newblock


\bibitem[Dong et~al\mbox{.}(2020)]%
        {dong2020survey}
\bibfield{author}{\bibinfo{person}{Xibin Dong}, \bibinfo{person}{Zhiwen Yu},
  \bibinfo{person}{Wenming Cao}, \bibinfo{person}{Yifan Shi}, {and}
  \bibinfo{person}{Qianli Ma}.} \bibinfo{year}{2020}\natexlab{}.
\newblock \showarticletitle{A survey on ensemble learning}.
\newblock \bibinfo{journal}{\emph{Frontiers of Computer Science}}
  \bibinfo{volume}{14} (\bibinfo{year}{2020}), \bibinfo{pages}{241--258}.
\newblock


\bibitem[Dong et~al\mbox{.}(2017)]%
        {dong2017metapath2vec}
\bibfield{author}{\bibinfo{person}{Yuxiao Dong}, \bibinfo{person}{Nitesh~V
  Chawla}, {and} \bibinfo{person}{Ananthram Swami}.}
  \bibinfo{year}{2017}\natexlab{}.
\newblock \showarticletitle{metapath2vec: Scalable representation learning for
  heterogeneous networks}. In \bibinfo{booktitle}{\emph{Proceedings of the 23rd
  ACM SIGKDD international conference on knowledge discovery and data mining}}.
  \bibinfo{pages}{135--144}.
\newblock


\bibitem[Fu et~al\mbox{.}(2016)]%
        {fu2016predicting}
\bibfield{author}{\bibinfo{person}{Gang Fu}, \bibinfo{person}{Ying Ding},
  \bibinfo{person}{Abhik Seal}, \bibinfo{person}{Bin Chen},
  \bibinfo{person}{Yizhou Sun}, {and} \bibinfo{person}{Evan Bolton}.}
  \bibinfo{year}{2016}\natexlab{}.
\newblock \showarticletitle{Predicting drug target interactions using
  meta-path-based semantic network analysis}.
\newblock \bibinfo{journal}{\emph{BMC bioinformatics}} \bibinfo{volume}{17},
  \bibinfo{number}{1} (\bibinfo{year}{2016}), \bibinfo{pages}{1--10}.
\newblock


\bibitem[Getoor et~al\mbox{.}(2001)]%
        {getoor2001learning}
\bibfield{author}{\bibinfo{person}{Lise Getoor}, \bibinfo{person}{Nir
  Friedman}, \bibinfo{person}{Daphne Koller}, {and} \bibinfo{person}{Avi
  Pfeffer}.} \bibinfo{year}{2001}\natexlab{}.
\newblock \showarticletitle{Learning probabilistic relational models}.
\newblock \bibinfo{journal}{\emph{Relational data mining}}
  (\bibinfo{year}{2001}), \bibinfo{pages}{307--335}.
\newblock


\bibitem[Grover and Leskovec(2016)]%
        {grover2016node2vec}
\bibfield{author}{\bibinfo{person}{Aditya Grover} {and} \bibinfo{person}{Jure
  Leskovec}.} \bibinfo{year}{2016}\natexlab{}.
\newblock \showarticletitle{node2vec: Scalable feature learning for networks}.
  In \bibinfo{booktitle}{\emph{Proceedings of the 22nd ACM SIGKDD international
  conference on Knowledge discovery and data mining}}.
  \bibinfo{pages}{855--864}.
\newblock


\bibitem[Hastie et~al\mbox{.}(2009)]%
        {hastie2009elements}
\bibfield{author}{\bibinfo{person}{Trevor Hastie}, \bibinfo{person}{Robert
  Tibshirani}, {and} \bibinfo{person}{Jerome~H Friedman}.}
  \bibinfo{year}{2009}\natexlab{}.
\newblock \bibinfo{booktitle}{\emph{The elements of statistical learning: data
  mining, inference, and prediction}}. Vol.~\bibinfo{volume}{2}.
\newblock \bibinfo{publisher}{Springer}.
\newblock


\bibitem[Hinton et~al\mbox{.}(2015)]%
        {hinton2015distilling}
\bibfield{author}{\bibinfo{person}{Geoffrey Hinton}, \bibinfo{person}{Oriol
  Vinyals}, {and} \bibinfo{person}{Jeff Dean}.}
  \bibinfo{year}{2015}\natexlab{}.
\newblock \showarticletitle{Distilling the knowledge in a neural network}.
\newblock \bibinfo{journal}{\emph{arXiv preprint arXiv:1503.02531}}
  (\bibinfo{year}{2015}).
\newblock


\bibitem[Hu and Lau(2013)]%
        {hu2013survey}
\bibfield{author}{\bibinfo{person}{Pili Hu} {and} \bibinfo{person}{Wing~Cheong
  Lau}.} \bibinfo{year}{2013}\natexlab{}.
\newblock \showarticletitle{A survey and taxonomy of graph sampling}.
\newblock \bibinfo{journal}{\emph{arXiv preprint arXiv:1308.5865}}
  (\bibinfo{year}{2013}).
\newblock


\bibitem[Hu et~al\mbox{.}(2020)]%
        {hu2020heterogeneous}
\bibfield{author}{\bibinfo{person}{Ziniu Hu}, \bibinfo{person}{Yuxiao Dong},
  \bibinfo{person}{Kuansan Wang}, {and} \bibinfo{person}{Yizhou Sun}.}
  \bibinfo{year}{2020}\natexlab{}.
\newblock \showarticletitle{Heterogeneous graph transformer}. In
  \bibinfo{booktitle}{\emph{Proceedings of the web conference 2020}}.
  \bibinfo{pages}{2704--2710}.
\newblock


\bibitem[Huang et~al\mbox{.}(2015)]%
        {huang2015social}
\bibfield{author}{\bibinfo{person}{Shangrong Huang}, \bibinfo{person}{Jian
  Zhang}, \bibinfo{person}{Lei Wang}, {and} \bibinfo{person}{Xian-Sheng Hua}.}
  \bibinfo{year}{2015}\natexlab{}.
\newblock \showarticletitle{Social friend recommendation based on multiple
  network correlation}.
\newblock \bibinfo{journal}{\emph{IEEE transactions on multimedia}}
  \bibinfo{volume}{18}, \bibinfo{number}{2} (\bibinfo{year}{2015}),
  \bibinfo{pages}{287--299}.
\newblock


\bibitem[Jaccard(1901)]%
        {jaccard1901etude}
\bibfield{author}{\bibinfo{person}{Paul Jaccard}.}
  \bibinfo{year}{1901}\natexlab{}.
\newblock \showarticletitle{{\'E}tude comparative de la distribution florale
  dans une portion des Alpes et des Jura}.
\newblock \bibinfo{journal}{\emph{Bull Soc Vaudoise Sci Nat}}
  \bibinfo{volume}{37} (\bibinfo{year}{1901}), \bibinfo{pages}{547--579}.
\newblock


\bibitem[Khosla et~al\mbox{.}(2020)]%
        {khosla2020supervised}
\bibfield{author}{\bibinfo{person}{Prannay Khosla}, \bibinfo{person}{Piotr
  Teterwak}, \bibinfo{person}{Chen Wang}, \bibinfo{person}{Aaron Sarna},
  \bibinfo{person}{Yonglong Tian}, \bibinfo{person}{Phillip Isola},
  \bibinfo{person}{Aaron Maschinot}, \bibinfo{person}{Ce Liu}, {and}
  \bibinfo{person}{Dilip Krishnan}.} \bibinfo{year}{2020}\natexlab{}.
\newblock \showarticletitle{Supervised contrastive learning}.
\newblock \bibinfo{journal}{\emph{Advances in neural information processing
  systems}}  \bibinfo{volume}{33} (\bibinfo{year}{2020}),
  \bibinfo{pages}{18661--18673}.
\newblock


\bibitem[Kreuzer et~al\mbox{.}(2021)]%
        {kreuzer2021rethinking}
\bibfield{author}{\bibinfo{person}{Devin Kreuzer}, \bibinfo{person}{Dominique
  Beaini}, \bibinfo{person}{Will Hamilton}, \bibinfo{person}{Vincent
  L{\'e}tourneau}, {and} \bibinfo{person}{Prudencio Tossou}.}
  \bibinfo{year}{2021}\natexlab{}.
\newblock \showarticletitle{Rethinking graph transformers with spectral
  attention}.
\newblock \bibinfo{journal}{\emph{Advances in Neural Information Processing
  Systems}}  \bibinfo{volume}{34} (\bibinfo{year}{2021}),
  \bibinfo{pages}{21618--21629}.
\newblock


\bibitem[Kumar et~al\mbox{.}(2020)]%
        {kumar2020link}
\bibfield{author}{\bibinfo{person}{Ajay Kumar},
  \bibinfo{person}{Shashank~Sheshar Singh}, \bibinfo{person}{Kuldeep Singh},
  {and} \bibinfo{person}{Bhaskar Biswas}.} \bibinfo{year}{2020}\natexlab{}.
\newblock \showarticletitle{Link prediction techniques, applications, and
  performance: A survey}.
\newblock \bibinfo{journal}{\emph{Physica A: Statistical Mechanics and its
  Applications}}  \bibinfo{volume}{553} (\bibinfo{year}{2020}),
  \bibinfo{pages}{124289}.
\newblock


\bibitem[Li et~al\mbox{.}(2023)]%
        {li2023metapath}
\bibfield{author}{\bibinfo{person}{Mei Li}, \bibinfo{person}{Xiangrui Cai},
  \bibinfo{person}{Sihan Xu}, {and} \bibinfo{person}{Hua Ji}.}
  \bibinfo{year}{2023}\natexlab{}.
\newblock \showarticletitle{Metapath-aggregated heterogeneous graph neural
  network for drug--target interaction prediction}.
\newblock \bibinfo{journal}{\emph{Briefings in Bioinformatics}}
  \bibinfo{volume}{24}, \bibinfo{number}{1} (\bibinfo{year}{2023}),
  \bibinfo{pages}{bbac578}.
\newblock


\bibitem[Li et~al\mbox{.}(2022)]%
        {li2022supervised}
\bibfield{author}{\bibinfo{person}{Yang Li}, \bibinfo{person}{Guanyu Qiao},
  \bibinfo{person}{Xin Gao}, {and} \bibinfo{person}{Guohua Wang}.}
  \bibinfo{year}{2022}\natexlab{}.
\newblock \showarticletitle{Supervised graph co-contrastive learning for
  drug--target interaction prediction}.
\newblock \bibinfo{journal}{\emph{Bioinformatics}} \bibinfo{volume}{38},
  \bibinfo{number}{10} (\bibinfo{year}{2022}), \bibinfo{pages}{2847--2854}.
\newblock


\bibitem[Lin et~al\mbox{.}(2020)]%
        {lin2020kgnn}
\bibfield{author}{\bibinfo{person}{Xuan Lin}, \bibinfo{person}{Zhe Quan},
  \bibinfo{person}{Zhi-Jie Wang}, \bibinfo{person}{Tengfei Ma}, {and}
  \bibinfo{person}{Xiangxiang Zeng}.} \bibinfo{year}{2020}\natexlab{}.
\newblock \showarticletitle{KGNN: Knowledge Graph Neural Network for Drug-Drug
  Interaction Prediction.}. In \bibinfo{booktitle}{\emph{IJCAI}},
  Vol.~\bibinfo{volume}{380}. \bibinfo{pages}{2739--2745}.
\newblock


\bibitem[Liu et~al\mbox{.}(2016)]%
        {liu2016neighborhood}
\bibfield{author}{\bibinfo{person}{Yong Liu}, \bibinfo{person}{Min Wu},
  \bibinfo{person}{Chunyan Miao}, \bibinfo{person}{Peilin Zhao}, {and}
  \bibinfo{person}{Xiao-Li Li}.} \bibinfo{year}{2016}\natexlab{}.
\newblock \showarticletitle{Neighborhood regularized logistic matrix
  factorization for drug-target interaction prediction}.
\newblock \bibinfo{journal}{\emph{PLoS computational biology}}
  \bibinfo{volume}{12}, \bibinfo{number}{2} (\bibinfo{year}{2016}),
  \bibinfo{pages}{e1004760}.
\newblock


\bibitem[Luo et~al\mbox{.}(2017)]%
        {luo2017network}
\bibfield{author}{\bibinfo{person}{Yunan Luo}, \bibinfo{person}{Xinbin Zhao},
  \bibinfo{person}{Jingtian Zhou}, \bibinfo{person}{Jinglin Yang},
  \bibinfo{person}{Yanqing Zhang}, \bibinfo{person}{Wenhua Kuang},
  \bibinfo{person}{Jian Peng}, \bibinfo{person}{Ligong Chen}, {and}
  \bibinfo{person}{Jianyang Zeng}.} \bibinfo{year}{2017}\natexlab{}.
\newblock \showarticletitle{A network integration approach for drug-target
  interaction prediction and computational drug repositioning from
  heterogeneous information}.
\newblock \bibinfo{journal}{\emph{Nature communications}} \bibinfo{volume}{8},
  \bibinfo{number}{1} (\bibinfo{year}{2017}), \bibinfo{pages}{573}.
\newblock


\bibitem[Marin and Wellman(2011)]%
        {marin2011social}
\bibfield{author}{\bibinfo{person}{Alexandra Marin} {and}
  \bibinfo{person}{Barry Wellman}.} \bibinfo{year}{2011}\natexlab{}.
\newblock \showarticletitle{Social network analysis: An introduction}.
\newblock \bibinfo{journal}{\emph{The SAGE handbook of social network
  analysis}}  \bibinfo{volume}{11} (\bibinfo{year}{2011}), \bibinfo{pages}{25}.
\newblock


\bibitem[M{\"u}ller et~al\mbox{.}(2023)]%
        {muller2023attending}
\bibfield{author}{\bibinfo{person}{Luis M{\"u}ller}, \bibinfo{person}{Mikhail
  Galkin}, \bibinfo{person}{Christopher Morris}, {and}
  \bibinfo{person}{Ladislav Ramp{\'a}{\v{s}}ek}.}
  \bibinfo{year}{2023}\natexlab{}.
\newblock \showarticletitle{Attending to graph transformers}.
\newblock \bibinfo{journal}{\emph{arXiv preprint arXiv:2302.04181}}
  (\bibinfo{year}{2023}).
\newblock


\bibitem[Peng et~al\mbox{.}(2021)]%
        {peng2021end}
\bibfield{author}{\bibinfo{person}{Jiajie Peng}, \bibinfo{person}{Yuxian Wang},
  \bibinfo{person}{Jiaojiao Guan}, \bibinfo{person}{Jingyi Li},
  \bibinfo{person}{Ruijiang Han}, \bibinfo{person}{Jianye Hao},
  \bibinfo{person}{Zhongyu Wei}, {and} \bibinfo{person}{Xuequn Shang}.}
  \bibinfo{year}{2021}\natexlab{}.
\newblock \showarticletitle{An end-to-end heterogeneous graph representation
  learning-based framework for drug--target interaction prediction}.
\newblock \bibinfo{journal}{\emph{Briefings in bioinformatics}}
  \bibinfo{volume}{22}, \bibinfo{number}{5} (\bibinfo{year}{2021}),
  \bibinfo{pages}{bbaa430}.
\newblock


\bibitem[Peng et~al\mbox{.}(2022)]%
        {peng2022smile}
\bibfield{author}{\bibinfo{person}{Miao Peng}, \bibinfo{person}{Ben Liu},
  \bibinfo{person}{Qianqian Xie}, \bibinfo{person}{Wenjie Xu},
  \bibinfo{person}{Hua Wang}, {and} \bibinfo{person}{Min Peng}.}
  \bibinfo{year}{2022}\natexlab{}.
\newblock \showarticletitle{SMiLE: Schema-augmented Multi-level Contrastive
  Learning for Knowledge Graph Link Prediction}.
\newblock \bibinfo{journal}{\emph{arXiv preprint arXiv:2210.04870}}
  (\bibinfo{year}{2022}).
\newblock


\bibitem[Perlman et~al\mbox{.}(2011)]%
        {perlman2011combining}
\bibfield{author}{\bibinfo{person}{Liat Perlman}, \bibinfo{person}{Assaf
  Gottlieb}, \bibinfo{person}{Nir Atias}, \bibinfo{person}{Eytan Ruppin}, {and}
  \bibinfo{person}{Roded Sharan}.} \bibinfo{year}{2011}\natexlab{}.
\newblock \showarticletitle{Combining drug and gene similarity measures for
  drug-target elucidation}.
\newblock \bibinfo{journal}{\emph{Journal of computational biology}}
  \bibinfo{volume}{18}, \bibinfo{number}{2} (\bibinfo{year}{2011}),
  \bibinfo{pages}{133--145}.
\newblock


\bibitem[Ramp{\'a}{\v{s}}ek et~al\mbox{.}(2022)]%
        {rampavsek2022recipe}
\bibfield{author}{\bibinfo{person}{Ladislav Ramp{\'a}{\v{s}}ek},
  \bibinfo{person}{Michael Galkin}, \bibinfo{person}{Vijay~Prakash Dwivedi},
  \bibinfo{person}{Anh~Tuan Luu}, \bibinfo{person}{Guy Wolf}, {and}
  \bibinfo{person}{Dominique Beaini}.} \bibinfo{year}{2022}\natexlab{}.
\newblock \showarticletitle{Recipe for a general, powerful, scalable graph
  transformer}.
\newblock \bibinfo{journal}{\emph{Advances in Neural Information Processing
  Systems}}  \bibinfo{volume}{35} (\bibinfo{year}{2022}),
  \bibinfo{pages}{14501--14515}.
\newblock


\bibitem[Sagi and Rokach(2018)]%
        {sagi2018ensemble}
\bibfield{author}{\bibinfo{person}{Omer Sagi} {and} \bibinfo{person}{Lior
  Rokach}.} \bibinfo{year}{2018}\natexlab{}.
\newblock \showarticletitle{Ensemble learning: A survey}.
\newblock \bibinfo{journal}{\emph{Wiley Interdisciplinary Reviews: Data Mining
  and Knowledge Discovery}} \bibinfo{volume}{8}, \bibinfo{number}{4}
  (\bibinfo{year}{2018}), \bibinfo{pages}{e1249}.
\newblock


\bibitem[Salton(1983)]%
        {salton1983introduction}
\bibfield{author}{\bibinfo{person}{Gerard Salton}.}
  \bibinfo{year}{1983}\natexlab{}.
\newblock \showarticletitle{Introduction to modern information retrieval}.
\newblock \bibinfo{journal}{\emph{McGraw-Hill}} (\bibinfo{year}{1983}).
\newblock


\bibitem[Udrescu et~al\mbox{.}(2016)]%
        {udrescu2016clustering}
\bibfield{author}{\bibinfo{person}{Lucre{\c{t}}ia Udrescu},
  \bibinfo{person}{Laura Sb{\^a}rcea}, \bibinfo{person}{Alexandru
  Top{\^\i}rceanu}, \bibinfo{person}{Alexandru Iovanovici},
  \bibinfo{person}{Ludovic Kurunczi}, \bibinfo{person}{Paul Bogdan}, {and}
  \bibinfo{person}{Mihai Udrescu}.} \bibinfo{year}{2016}\natexlab{}.
\newblock \showarticletitle{Clustering drug-drug interaction networks with
  energy model layouts: community analysis and drug repurposing}.
\newblock \bibinfo{journal}{\emph{Scientific reports}} \bibinfo{volume}{6},
  \bibinfo{number}{1} (\bibinfo{year}{2016}), \bibinfo{pages}{32745}.
\newblock


\bibitem[Vaswani et~al\mbox{.}(2017)]%
        {vaswani2017attention}
\bibfield{author}{\bibinfo{person}{Ashish Vaswani}, \bibinfo{person}{Noam
  Shazeer}, \bibinfo{person}{Niki Parmar}, \bibinfo{person}{Jakob Uszkoreit},
  \bibinfo{person}{Llion Jones}, \bibinfo{person}{Aidan~N Gomez},
  \bibinfo{person}{{\L}ukasz Kaiser}, {and} \bibinfo{person}{Illia
  Polosukhin}.} \bibinfo{year}{2017}\natexlab{}.
\newblock \showarticletitle{Attention is all you need}.
\newblock \bibinfo{journal}{\emph{Advances in neural information processing
  systems}}  \bibinfo{volume}{30} (\bibinfo{year}{2017}).
\newblock


\bibitem[Wang et~al\mbox{.}(2019)]%
        {wang2019heterogeneous}
\bibfield{author}{\bibinfo{person}{Xiao Wang}, \bibinfo{person}{Houye Ji},
  \bibinfo{person}{Chuan Shi}, \bibinfo{person}{Bai Wang},
  \bibinfo{person}{Yanfang Ye}, \bibinfo{person}{Peng Cui}, {and}
  \bibinfo{person}{Philip~S Yu}.} \bibinfo{year}{2019}\natexlab{}.
\newblock \showarticletitle{Heterogeneous graph attention network}. In
  \bibinfo{booktitle}{\emph{WWW}}. \bibinfo{pages}{2022--2032}.
\newblock


\bibitem[Wen et~al\mbox{.}(2017)]%
        {wen2017deep}
\bibfield{author}{\bibinfo{person}{Ming Wen}, \bibinfo{person}{Zhimin Zhang},
  \bibinfo{person}{Shaoyu Niu}, \bibinfo{person}{Haozhi Sha},
  \bibinfo{person}{Ruihan Yang}, \bibinfo{person}{Yonghuan Yun}, {and}
  \bibinfo{person}{Hongmei Lu}.} \bibinfo{year}{2017}\natexlab{}.
\newblock \showarticletitle{Deep-learning-based drug--target interaction
  prediction}.
\newblock \bibinfo{journal}{\emph{Journal of proteome research}}
  \bibinfo{volume}{16}, \bibinfo{number}{4} (\bibinfo{year}{2017}),
  \bibinfo{pages}{1401--1409}.
\newblock


\bibitem[Xu et~al\mbox{.}(2022)]%
        {xu2022sociolink}
\bibfield{author}{\bibinfo{person}{Ruiyun~Rayna Xu}, \bibinfo{person}{Hailiang
  Chen}, {and} \bibinfo{person}{J~Leon Zhao}.} \bibinfo{year}{2022}\natexlab{}.
\newblock \showarticletitle{SocioLink: Leveraging Relational Information in
  Knowledge Graphs for Startup Recommendations}.
\newblock \bibinfo{journal}{\emph{Journal of Management Information Systems
  forthcoming}} (\bibinfo{year}{2022}).
\newblock


\bibitem[Ying et~al\mbox{.}(2021)]%
        {ying2021transformers}
\bibfield{author}{\bibinfo{person}{Chengxuan Ying}, \bibinfo{person}{Tianle
  Cai}, \bibinfo{person}{Shengjie Luo}, \bibinfo{person}{Shuxin Zheng},
  \bibinfo{person}{Guolin Ke}, \bibinfo{person}{Di He},
  \bibinfo{person}{Yanming Shen}, {and} \bibinfo{person}{Tie-Yan Liu}.}
  \bibinfo{year}{2021}\natexlab{}.
\newblock \showarticletitle{Do transformers really perform badly for graph
  representation?}
\newblock \bibinfo{journal}{\emph{Advances in Neural Information Processing
  Systems}}  \bibinfo{volume}{34} (\bibinfo{year}{2021}),
  \bibinfo{pages}{28877--28888}.
\newblock


\bibitem[You et~al\mbox{.}(2020)]%
        {you2020graph}
\bibfield{author}{\bibinfo{person}{Yuning You}, \bibinfo{person}{Tianlong
  Chen}, \bibinfo{person}{Yongduo Sui}, \bibinfo{person}{Ting Chen},
  \bibinfo{person}{Zhangyang Wang}, {and} \bibinfo{person}{Yang Shen}.}
  \bibinfo{year}{2020}\natexlab{}.
\newblock \showarticletitle{Graph contrastive learning with augmentations}.
\newblock \bibinfo{journal}{\emph{Advances in neural information processing
  systems}}  \bibinfo{volume}{33} (\bibinfo{year}{2020}),
  \bibinfo{pages}{5812--5823}.
\newblock


\bibitem[Zhang et~al\mbox{.}(2019)]%
        {zhang2019heterogeneous}
\bibfield{author}{\bibinfo{person}{Chuxu Zhang}, \bibinfo{person}{Dongjin
  Song}, \bibinfo{person}{Chao Huang}, \bibinfo{person}{Ananthram Swami}, {and}
  \bibinfo{person}{Nitesh~V Chawla}.} \bibinfo{year}{2019}\natexlab{}.
\newblock \showarticletitle{Heterogeneous graph neural network}. In
  \bibinfo{booktitle}{\emph{Proceedings of the 25th ACM SIGKDD international
  conference on knowledge discovery \& data mining}}.
  \bibinfo{pages}{793--803}.
\newblock


\bibitem[Zhang et~al\mbox{.}(2020)]%
        {zhang2020revisiting}
\bibfield{author}{\bibinfo{person}{Muhan Zhang}, \bibinfo{person}{Pan Li},
  \bibinfo{person}{Yinglong Xia}, \bibinfo{person}{Kai Wang}, {and}
  \bibinfo{person}{Long Jin}.} \bibinfo{year}{2020}\natexlab{}.
\newblock \showarticletitle{Revisiting graph neural networks for link
  prediction}.
\newblock  (\bibinfo{year}{2020}).
\newblock


\bibitem[Zhang et~al\mbox{.}(2022b)]%
        {zhang2022cat}
\bibfield{author}{\bibinfo{person}{Shengming Zhang}, \bibinfo{person}{Yanchi
  Liu}, \bibinfo{person}{Xuchao Zhang}, \bibinfo{person}{Wei Cheng},
  \bibinfo{person}{Haifeng Chen}, {and} \bibinfo{person}{Hui Xiong}.}
  \bibinfo{year}{2022}\natexlab{b}.
\newblock \showarticletitle{CAT: Beyond Efficient Transformer for Content-Aware
  Anomaly Detection in Event Sequences}. In
  \bibinfo{booktitle}{\emph{Proceedings of the 28th ACM SIGKDD Conference on
  Knowledge Discovery and Data Mining}}. \bibinfo{pages}{4541--4550}.
\newblock


\bibitem[Zhang and Sun(2023)]%
        {zhang2023meta}
\bibfield{author}{\bibinfo{person}{Shengming Zhang} {and}
  \bibinfo{person}{Yizhou Sun}.} \bibinfo{year}{2023}\natexlab{}.
\newblock \showarticletitle{Meta-Path-based Probabilistic Soft Logic for
  Drug-Target Interaction Prediction}.
\newblock \bibinfo{journal}{\emph{arXiv preprint arXiv:2306.13770}}
  (\bibinfo{year}{2023}).
\newblock


\bibitem[Zhang et~al\mbox{.}(2021)]%
        {zhang2021scalable}
\bibfield{author}{\bibinfo{person}{Shengming Zhang}, \bibinfo{person}{Hao
  Zhong}, \bibinfo{person}{Zixuan Yuan}, {and} \bibinfo{person}{Hui Xiong}.}
  \bibinfo{year}{2021}\natexlab{}.
\newblock \showarticletitle{Scalable heterogeneous graph neural networks for
  predicting high-potential early-stage startups}. In
  \bibinfo{booktitle}{\emph{Proceedings of the 27th ACM SIGKDD Conference on
  Knowledge Discovery \& Data Mining}}. \bibinfo{pages}{2202--2211}.
\newblock


\bibitem[Zhang et~al\mbox{.}(2022a)]%
        {zhang2022hierarchical}
\bibfield{author}{\bibinfo{person}{Zaixi Zhang}, \bibinfo{person}{Qi Liu},
  \bibinfo{person}{Qingyong Hu}, {and} \bibinfo{person}{Chee-Kong Lee}.}
  \bibinfo{year}{2022}\natexlab{a}.
\newblock \showarticletitle{Hierarchical graph transformer with adaptive node
  sampling}.
\newblock \bibinfo{journal}{\emph{Advances in Neural Information Processing
  Systems}}  \bibinfo{volume}{35} (\bibinfo{year}{2022}),
  \bibinfo{pages}{21171--21183}.
\newblock


\bibitem[Zhang et~al\mbox{.}(2023)]%
        {zhang2023line}
\bibfield{author}{\bibinfo{person}{Zehua Zhang}, \bibinfo{person}{Shilin Sun},
  \bibinfo{person}{Guixiang Ma}, {and} \bibinfo{person}{Caiming Zhong}.}
  \bibinfo{year}{2023}\natexlab{}.
\newblock \showarticletitle{Line graph contrastive learning for link
  prediction}.
\newblock \bibinfo{journal}{\emph{Pattern Recognition}}  \bibinfo{volume}{140}
  (\bibinfo{year}{2023}), \bibinfo{pages}{109537}.
\newblock


\bibitem[Zhou et~al\mbox{.}(2019)]%
        {zhou2019deep}
\bibfield{author}{\bibinfo{person}{Guorui Zhou}, \bibinfo{person}{Na Mou},
  \bibinfo{person}{Ying Fan}, \bibinfo{person}{Qi Pi}, \bibinfo{person}{Weijie
  Bian}, \bibinfo{person}{Chang Zhou}, \bibinfo{person}{Xiaoqiang Zhu}, {and}
  \bibinfo{person}{Kun Gai}.} \bibinfo{year}{2019}\natexlab{}.
\newblock \showarticletitle{Deep interest evolution network for click-through
  rate prediction}. In \bibinfo{booktitle}{\emph{Proceedings of the AAAI
  conference on artificial intelligence}}, Vol.~\bibinfo{volume}{33}.
  \bibinfo{pages}{5941--5948}.
\newblock


\bibitem[Zhu et~al\mbox{.}(2021)]%
        {zhu2021neural}
\bibfield{author}{\bibinfo{person}{Zhaocheng Zhu}, \bibinfo{person}{Zuobai
  Zhang}, \bibinfo{person}{Louis-Pascal Xhonneux}, {and} \bibinfo{person}{Jian
  Tang}.} \bibinfo{year}{2021}\natexlab{}.
\newblock \showarticletitle{Neural bellman-ford networks: A general graph
  neural network framework for link prediction}.
\newblock \bibinfo{journal}{\emph{Advances in Neural Information Processing
  Systems}} (\bibinfo{year}{2021}).
\newblock


\end{thebibliography}
